\documentclass[aps,pre,showpacs,superscriptaddress]{revtex4}
\usepackage{graphicx}
\usepackage{natbib}

\def\Mn{\hbox{Mn$_{12}$}}
\def\V{\hbox{V$_{15}$}}

\def\beq{\begin{equation}}
\def\eeq{\end{equation}}

\begin{document}
\title{Microscopic origin of the adiabatic change of magnetization in molecular magnets}
\author{H. De Raedt}
 \email{deraedt@phys.rug.nl}
 \homepage{http://www.compphys.org}
\affiliation{Department of Applied Physics-Computational Physics,
Materials Science Centre, University of Groningen, Nijenborgh 4,
NL-9747 AG Groningen, The Netherlands}
\author{S. Miyashita}
 \email{miya@spin.t.u-tokyo.ac.jp}
\affiliation{Department of Applied Physics, Graduate School of Engineering,
University of Tokyo, Bunkyo-ku Tokyo 113-8656, Japan}
\author{K. Michielsen}
 \email{kristel@phys.rug.nl}
\affiliation{Department of Applied Physics-Computational Physics,
Materials Science Centre, University of Groningen, Nijenborgh 4,
NL-9747 AG Groningen, The Netherlands}
\author{M. Machida}
 \email{machida@spin.t.u-tokyo.ac.jp}
\affiliation{Department of Applied Physics, Graduate School of Engineering,
University of Tokyo, Bunkyo-ku Tokyo 113-8656, Japan}
\begin{abstract}
A microscopic model of the molecular magnet \V\ is used to
study mechanisms for the adiabatic change of the magnetization
in time-dependent magnetic fields.
Effects of the Dzyaloshinskii-Moriya interaction,
the most plausible source for the energy-level repulsions
that lead to adiabatic changes of the magnetization,
are studied in detail.
We find that the energy-level repulsions that result from
this interaction exhibit a strong dependence on
the direction of the applied field.
We also discuss the role of magnetic anisotropy in the molecule \Mn -acetate.
\end{abstract}
\date{\today}

\maketitle

\section{Introduction}\label{sec1}

Recently, magnetic molecules such as \Mn\ or \V\ have attracted a lot of interest.
These nanomagnets are often used to study explicit real-time quantum dynamics,
e.g. tunneling of the magnetization and quantum (de)coherence~\cite{%
Gunter,%
Caneschi,%
gat,%
gat2,%
Levine,%
Friedman,%
Thomas,%
sangregorio,%
fe8tuna,%
fe8tunb,%
Bernard,%
Perenboom,%
Irinel3,%
Pohjola,%
Zhong,%
Irinel1,%
Irinel2,%
Bouk0,%
Bouk1,%
Wernsdorfer,%
Honecker,%
Irinel4%
}.
As a result of the very weak intermolecular interactions,
experiments can directly probe the magnetization dynamics of the individual molecules.
In particular, the adiabatic change of the magnetization at low-temperature
is governed by the discrete energy-level structure~\cite{Seiji0a,Seiji0b,Slava0,Gunter0,Hans2}.

The adiabatic change of the magnetization requires some interactions that
yield energy level repulsions, i.e. interactions that do not commute with the magnetization.
The Dzyaloshinskii-Moriya (DM) interaction is the most likely candidate
for such an interaction~\cite{Bernard,Miya-Naga,Konst}.
In the case of anisotropic high-spin molecules such as Mn$_{12}$ and Fe$_8$,
simplified anisotropic single-spin models for a specific spin multiplet
can approximately reproduce the gaps of the level repulsions.
However, the case of V$_{15}$ is more complicated
because V$_{15}$ has half-odd-integer spin
and the time-reversal symmetry enforces at least a two-fold degeneracy
of the energy levels at zero field.
%
%
%

It has been pointed out that the DM interaction is accompanied by a
higher-order correction term that restores the SU(2) symmetry~\cite{Kaplan,Shekntman1,Shekntman2,Zheludev}.
In this paper, we focus on the effects of the DM interaction and leave the inclusion of
the higher-order term for future study.
As another source of level repulsion, we might consider
the hyperfine interaction with the nuclear spin.
The effects of the hyperfine interaction have already been discussed in Ref.~\cite{HYPF}.

As the DM interaction has a vector character and is anisotropic,
the dynamics of the magnetization is expected to depend on the direction of the magnetic field.
First we study the characteristic properties of the DM interaction for
a simplified model of \V, namely three spins on a triangle.
Then we confirm the properties found in the three-spin model by full
diagonalization of the 15-spin model of \V.
In order to bridge the energy scales involved (e.g. from 800K, a typical energy scale for the
interaction between individual magnetic ions, to about $10^{-2}$K, a typical energy scale for
energy-level splittings),
the calculation of the energy levels of the many-spin Hamiltonian has to be very accurate.
We have tested various standard algorithms to compute the low-lying states.
For systems that are too large to be solved by full exact diagonalization
(such as the 15-spin \V\ model), we use the Lanczos method with full orthogonalization (LFO),
a Chebyshev polynomial projector (CPP) method,
and a power method with additional subspace diagonalization.
These algorithms can solve the rather large eigenvalue problems with sufficient accuracy.
The consistency of the data obtained by different methods gives
extra confidence in the numerical results.

The magnetic properties of molecules such as \Mn\ are often studied by considering
a simplified model for the magnetic energy levels for a specific spin multiplet,
e.g. $S=10$.
However, for these and other, similar, magnetic molecules that consist of several magnetic moments
(in the case of \Mn: eight Mn$^{3+}$($S=2$) and four Mn$^{4+}$($S=3/2$)), the reduction of the many-body Hamiltonian
to an effective Hamiltonian for a specific spin multiplet is non-trivial.
Magnetic anisotropy, a result of the geometrical arrangement of
the magnetic ions within a molecule of low symmetry, mixes states of different total spin
and enforces a treatment of the full Hilbert space of the system.
For Mn$_{12}$, the dominant contribution to the magnetic mixing
due to spin-orbit interactions is also given by
the DM interaction~\cite{Dzy,Mor}.
In principle, this type of interaction can change energy-level crossings into energy-level repulsions.
The presence of the latter is essential to explain
the adiabatic changes of the magnetization at the resonant fields~\cite{Seiji0a,Seiji0b,Slava0,Gunter0,Hans2}.
Thus, a minimal magnetic model Hamiltonian should contain (strong) Heisenberg interactions,
DM interactions and a coupling to the applied magnetic field~\cite{Bernard,
Misha1,%
mn12spl,%
RudraMn12,%
Rudra,%
Miya-Naga,%
Raghu,%
Hans1,%
Konst,%
RudraSeiji}.
Experiments on \Mn\ suggest that the energy gaps related to the transition from
a state with magnetization $M\approx-10$ to a state with magnetization
$M< 4$ are of the order of $10^{-9}$K~\cite{private}.
Such gaps are too small to detect with standard precision (13-14 digits) calculations
and therefore, in this paper we only present the global energy level diagram
obtained from microscopic model calculations.

The paper is organized as follows.
In Sec.~\ref{sec2} we analyze a reduced three-spin model with C$_3$ symmetry for the \V\ molecule.
Results for the energy level schemes for a 15-spin model of the \V\ are presented in Sec.~\ref{sec3}.
In Sec.~\ref{sec4} we discuss the effects of anisotropic terms and the cases with less symmetry.
In Sec.~\ref{sec5}, we report results for Mn$_{12}$. In Sec.~\ref{sec6} we give our conclusions.
In Appendix A and B, we briefly discuss some analytical solutions and
the numerical algorithms that we use to compute the energy levels, respectively.

\section{Triangle-model with C$_3$ symmetry}
\label{sec2}

As a simplified model for the \V\ molecule
we consider a system of three spins on a triangle with
C$_3$ symmetry.
We choose the $z$-axis to lie along the axis of C$_3$ symmetry.
The Hamiltonian is given by

\begin{eqnarray}
{\cal H}&=& -\sum_{i=1}^3 J_{i,i+1}{\bf S}_i \cdot{\bf S}_{i+1}
  + \sum_{\langle i,j\rangle} {\bf D}_{i,j}\cdot[{\bf S}_i\times {\bf S}_j]
     - {\bf h}\cdot\left(\sum_{i} {\bf S}_{i}\right),
\label{HD}
\label{VHam}
\end{eqnarray}
where $J_{1,2}=J_{2,3}=J_{3,1}\equiv J$, denotes the exchange interaction and
${\bf h}$ represents the applied magnetic field.
In general we can choose any direction of
the DM vector ${\bf D}_{i,j}$
unless there is some additional symmetry.
In the present case, because of the C$_3$ symmetry,
the $z$-component of the DM vectors must all be equal, i.e.
\begin{equation}
D_{1,2}^z=D_{2,3}^z=D_{3,1}^z\equiv D_z,
\label{d3sym}
\end{equation}
and the $x$ and $y$-components of the DM vector, $D_{i,j}^x$, $D_{i,j}^y$, must obey the relation

\begin{equation}
\left(\begin{array}{c} D_{2,3}^x \\ D_{2,3}^y \end{array} \right)
=
\left(\begin{array}{cc} -{1\over2} & {\sqrt{3}\over2} \\ -{\sqrt{3}\over2} & -{1\over2}  \end{array} \right)
\left(\begin{array}{c} D_x \\ D_y \end{array} \right),
\quad
\left(\begin{array}{c} D_{3,1}^x \\ D_{3,1}^y \end{array} \right)
=
\left(\begin{array}{cc} -{1\over2} & -{\sqrt{3}\over2} \\ {\sqrt{3}\over2} & -{1\over2}  \end{array} \right)
\left(\begin{array}{c} D_x \\ D_y \end{array} \right),
\label{c3sym}
\end{equation}
where $D_x\equiv D_{1,2}^x$ and $D_y\equiv D_{1,2}^y$.

In Appendix A we give the analytic expressions for the eigenvalues and
eigenvectors of model (\ref{HD})
in the case that the applied field is parallel to the $z$-axis and
the conditions (\ref{d3sym}) and (\ref{c3sym}) hold.
Then, the Hamiltonian (\ref{HD}) is block diagonal,
the matrix containing 4 blocks of $2\times2$ matrices.
The (8-dimensional) Hilbert space
separates into 4 two-dimensional spaces

\begin{equation}
\{|3/2,3/2\rangle,|a\rangle\}\quad,\quad
\{|3/2,1/2\rangle,|\bar a\rangle\}\quad,\quad
\{|3/2,-1/2\rangle,|\bar b\rangle\}\quad,\quad
\{|3/2,-3/2\rangle,|b\rangle\},
\label{states}
\end{equation}
where the state $|S,M\rangle$ denotes an eigenstate of the Heisenberg model
with total spin $S$ and magnetization $M$.
The expressions for the orthonormal states
$|a\rangle$ and $|b\rangle$ are given in appendix A.
The state $|\bar a\rangle$ ($|\bar b\rangle$)
denotes the state $|a\rangle$ ($|b\rangle$) with all spins reversed.
From Eq.~(\ref{states}) it follows that
there is no mixing among the four levels with $S=\pm 1/2$,
and therefore they simply cross each other.
From the analytical solution it is also easy to see that
there are no energy level repulsions at $h=0$.
Furthermore one can show analytically
that it is impossible to change the magnetization from $-3/2$ to $+3/2$
by (adiabatically) reversing the external field along the $z$-axis.

\begin{figure}[t]
\begin{center}
\setlength{\unitlength}{1cm}
\begin{picture}(14,6)
\put(-2.,0.){\includegraphics[width=9cm]{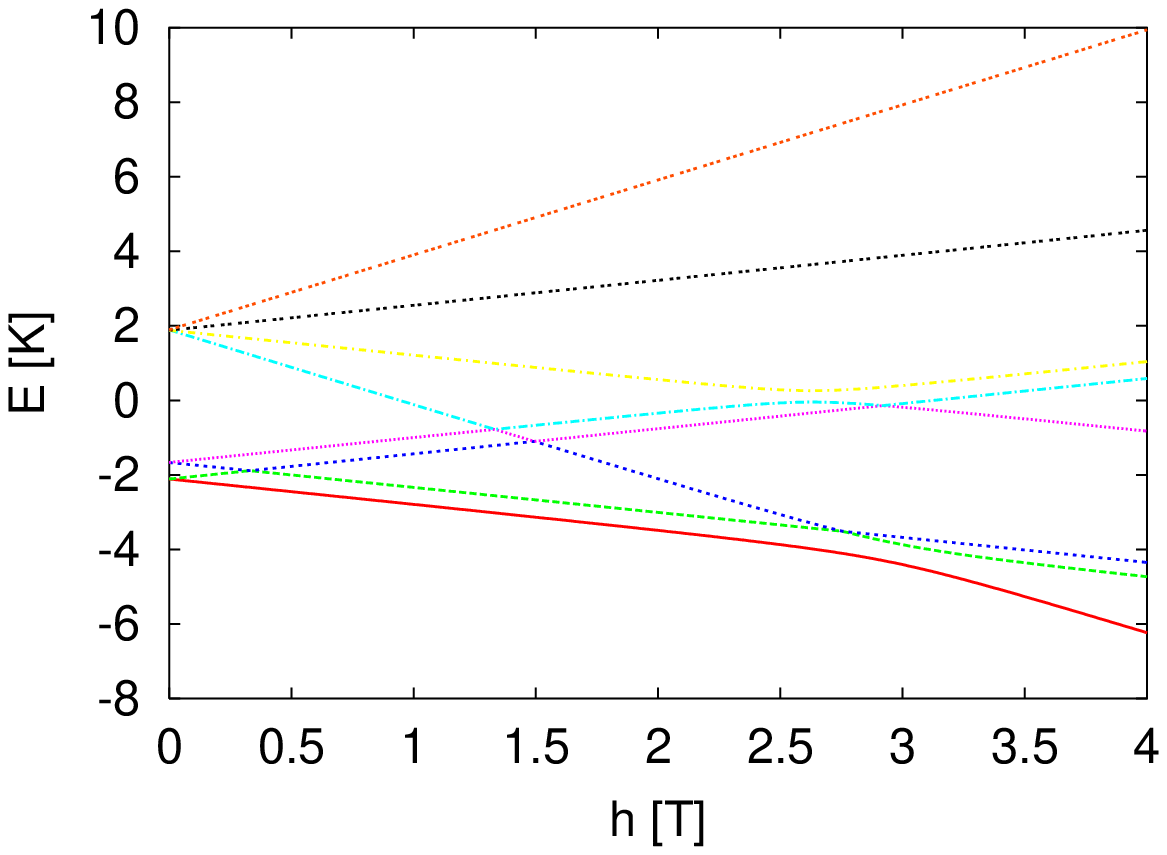}}
\put(7.,0.){\includegraphics[width=9cm]{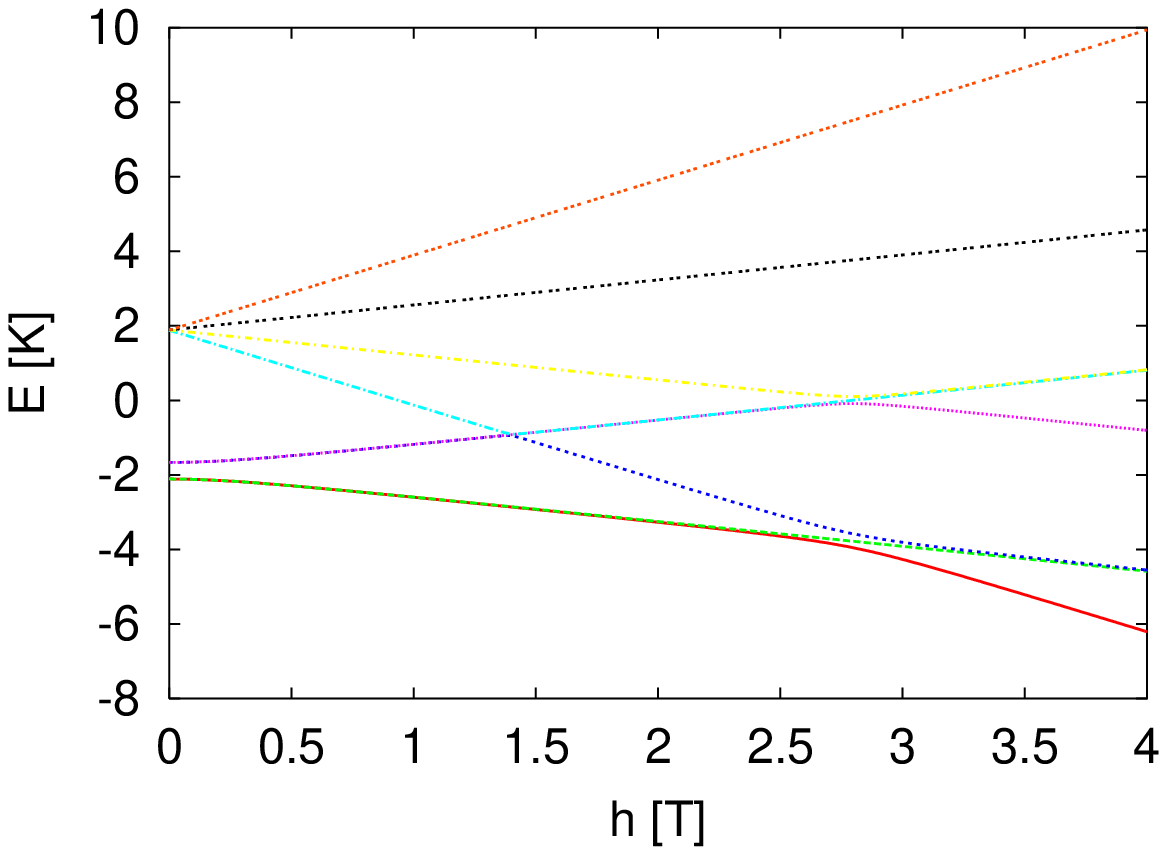}}
\end{picture}
\caption{%
Energy levels of Hamiltonian (\ref{VHam}) for $J=-2.5$K,
$D_x=D_y=D_z=0.25$K.
Left: Applied magnetic field ${\bf h}$ parallel
to the $z$-axis.
Right:
Applied magnetic field ${\bf h}$ along the $x$-axis.
}
\label{fig1}
\label{fig2}
\end{center}
\end{figure}

\begin{figure}[t]
\begin{center}
\setlength{\unitlength}{1cm}
\begin{picture}(14,6)
\put(-2.,0.){\includegraphics[width=9cm]{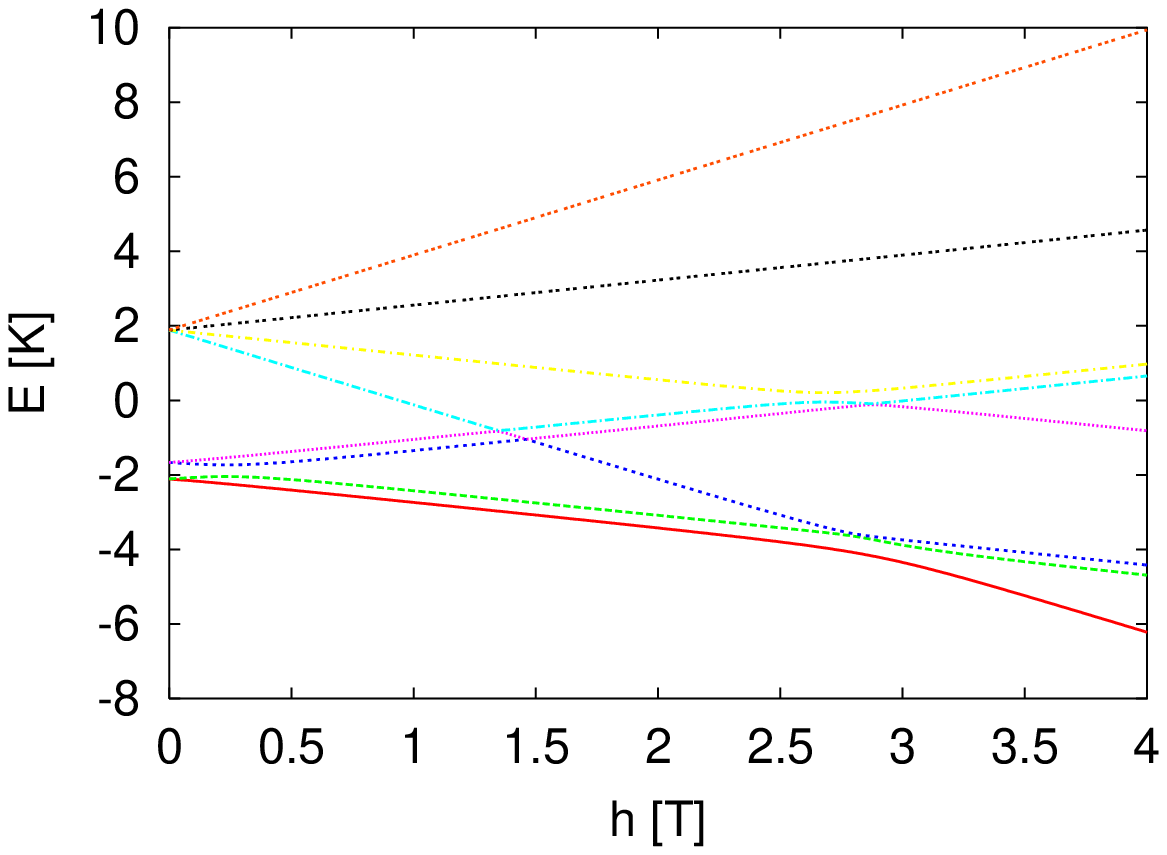}}
\put(7.,0.){\includegraphics[width=9cm]{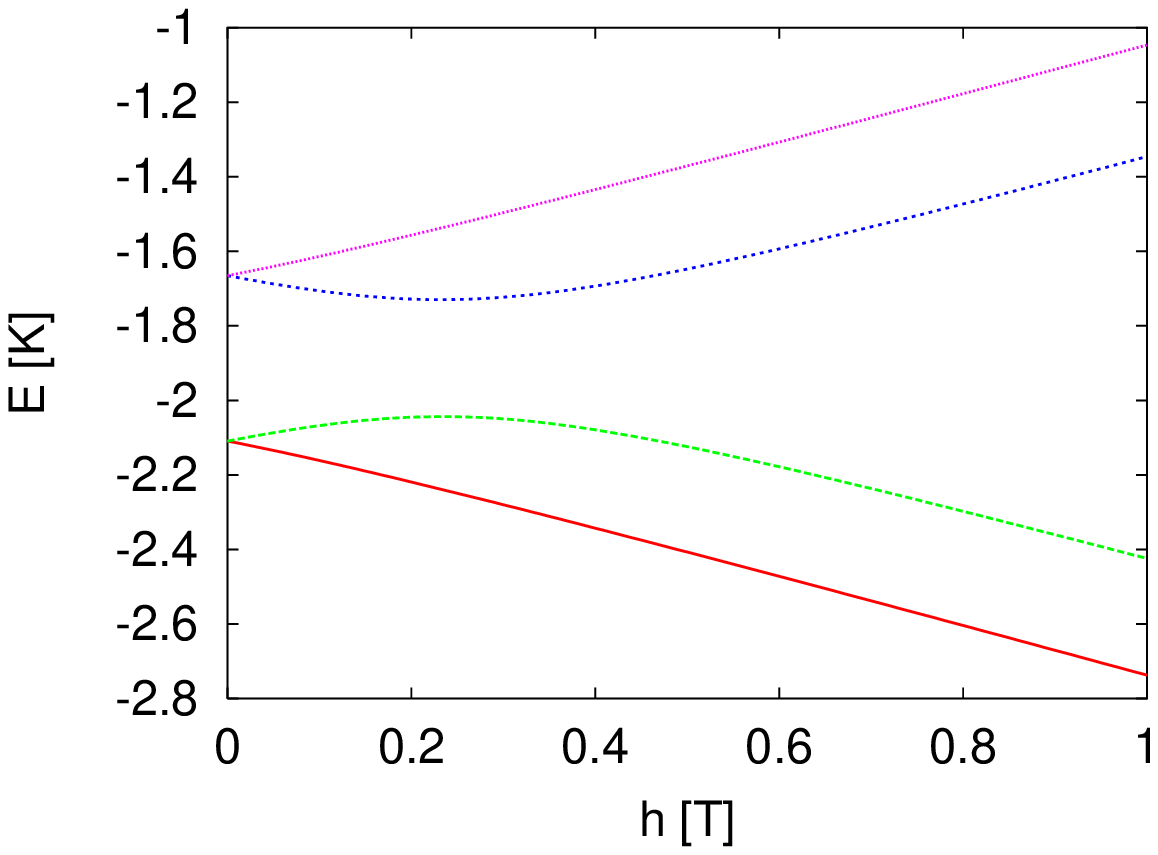}}
\end{picture}
\caption{%
Left:
Energy levels of Hamiltonian (\ref{VHam}) for $J=-2.5$K,
$D_x=D_y=D_z=0.25$K and the applied magnetic field ${\bf h}=h(1,0,1)/\sqrt{2}$
tilted by 45$^\circ$ with respect to the $z$-axis.
Right:
Detailed view of the $h$-dependence of the four lowest energy levels.
}
\label{fig3}
\label{fig4}
\end{center}
\end{figure}

In the left panel of Fig.~\ref{fig1} we show the energy levels as a function of
the strength $h$ of the applied magnetic field for the case $D_x=D_y=D_z$. The field is aligned along the $z$-axis.
It is evident that there is no level mixing at $h=0$.
It is important to note that the small energy difference between the first excited state and the degenerate ground
states at $h=0$ is not a tunneling gap.
Therefore the magnetization
does not change if the field ${\bf h}$ changes sign.
On the other hand, if we apply the magnetization along the $x$-direction
(i.e. perpendicular to the symmetry axis of the triangle),
then two of the states become degenerate.
We find two nearly degenerate avoided level crossings, as shown in the right panel of Fig.~\ref{fig1}.
As an intermediate case, in Fig.~\ref{fig4}
we show the energy level diagram for the case that the magnetic field is
tilted by 45$^{\circ}$ with respect to the $z$-axis (${\bf h}=h(1,0,1)/\sqrt{2}$).
Then, there is a simple crossing at zero field and an avoided level crossing
between the levels of $M=\pm 1/2$ at a finite magnetic field.
Indeed, a closer look at the level diagram
(see right panel in Fig.~\ref{fig4}) reveals that the
minimum energy difference between the two pairs of levels does not occur at
zero field but at $h\approx 0.35$T.
This implies that the Landau-Zener-St\"uckelberg transition
from the $|1/2,-1/2\rangle$ to the $|1/2,1/2\rangle$ level does not
take place at $h=0$ but at $h\approx 0.35$T.
In conclusion, the position and the energy splitting of the avoided
level crossing that is responsible for the adiabatic change
depend on the direction of the field.

The numerical results discussed above have been obtained for $D_x=D_y=D_z$.
In Ref.~\cite{Irinel4} the DM vector is taken parallel
to the $y$-axis at all the bonds and the field is applied along the $z$-axis.
For the present model, this case corresponds to the case
with only $D_z$ ($D_x=D_y=0$) and a field applied in the $x$-direction.
In this case, the gap opens symmetrically with respect to the applied field~\cite{Irinel4}.
However, Ref.~\cite{Irinel4} did not address the dependence on the direction of
the magnetic field.
In the next section we repeat the analysis of this dependence for a 15-spin model of the \V\ molecule.

\section{15-spin model for the Vanadium complex \V}
\label{sec3}

\begin{figure}[t]
\begin{center}
\includegraphics[width=6cm]{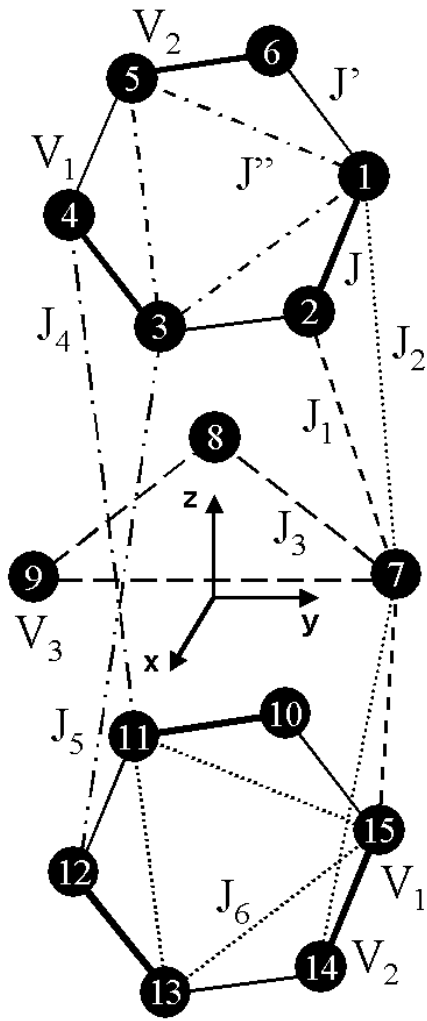}
\caption{%
Schematic diagram of the magnetic interactions in model (\ref{VHam}) of the \V\ molecule.}
\label{fig5}
\end{center}
\end{figure}

\subsection{Spin interactions in \V}

In Fig.~\ref{fig5}, we show the schematic diagram of the dominant magnetic (Heisenberg) interactions
in the 15-spin model of
the \V\ molecule
(K$_{6}$[V$^{IV}_{15}$As$_{6}$O$_{42}$(H$_{2}$O)]$\cdot$8H$_{2}$O).
The magnetic structure consists of two hexagons with six $S=1/2$ spins each, enclosing a triangle with
three $S=1/2$ spins. All dominant Heisenberg interactions are antiferromagnetic.
The dimension of the Hilbert space of this model is $2^{15}=32768$.
The minimal Hamiltonian is given by expression
(\ref{HD}) with 15 instead of 3 spins~\cite{Miya-Naga,Rudra,Konst,Irinel4}.
The Heisenberg interactions $J_{i,j}$ in Eq.~(\ref{HD}) between the vanadium atoms are defined according to Fig.~\ref{fig5}.
For simplicity, we assume that ${\bf D}_{i,j}=0$
except for bonds for which the Heisenberg exchange constant is the largest (i.e. equal to $J$ )~\cite{Konst,Rudra}.
Rotations about $2\pi/3$ and $4\pi/3$ around the axis perpendicular
to and passing through the center of the
hexagons leave the \V\ complex invariant.
This enforces the constraints (\ref{d3sym}) and (\ref{c3sym}) on the values of ${\bf D}_{i,j}$~\cite{Konst,RudraSeiji}.

\subsection{Energy level diagrams}

\begin{figure}[t]
\begin{center}
\setlength{\unitlength}{1cm}
\begin{picture}(14,6)
\put(-2.,0.){\includegraphics[width=9cm]{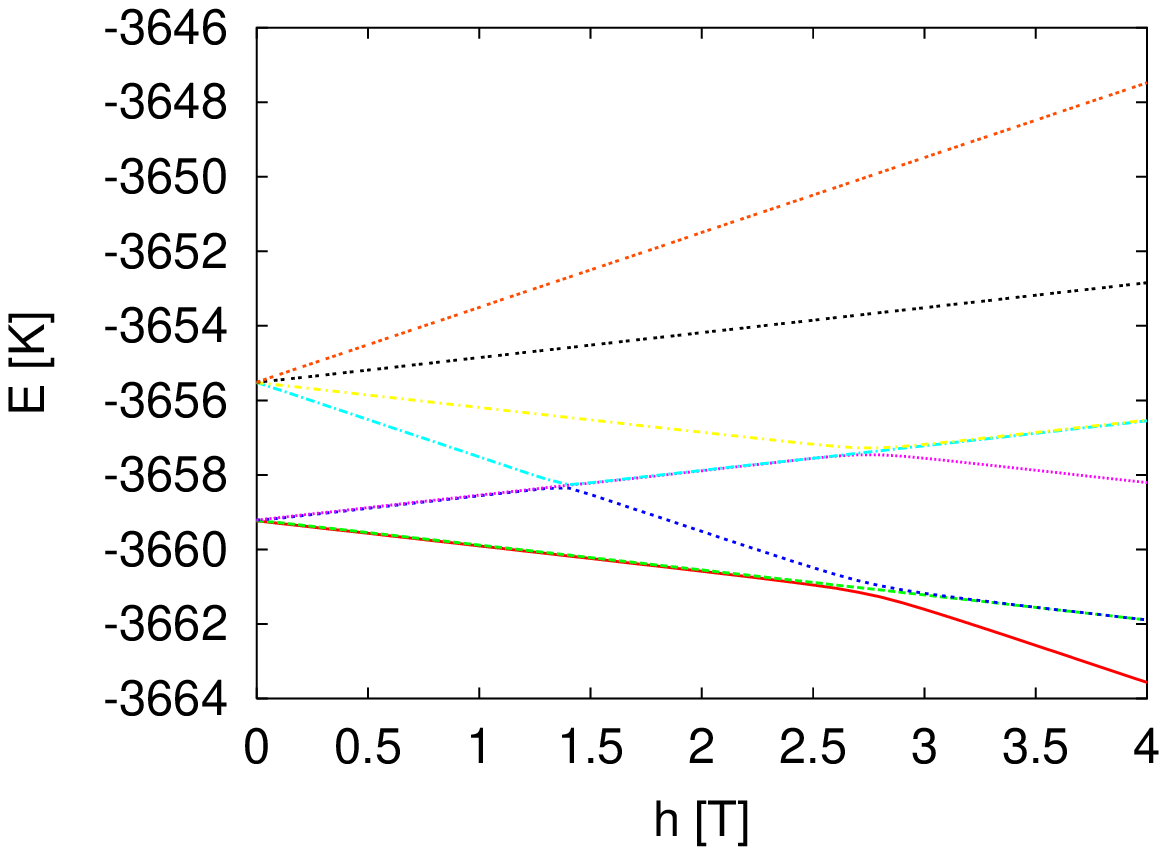}}
\put(7.,0.){\includegraphics[width=9cm]{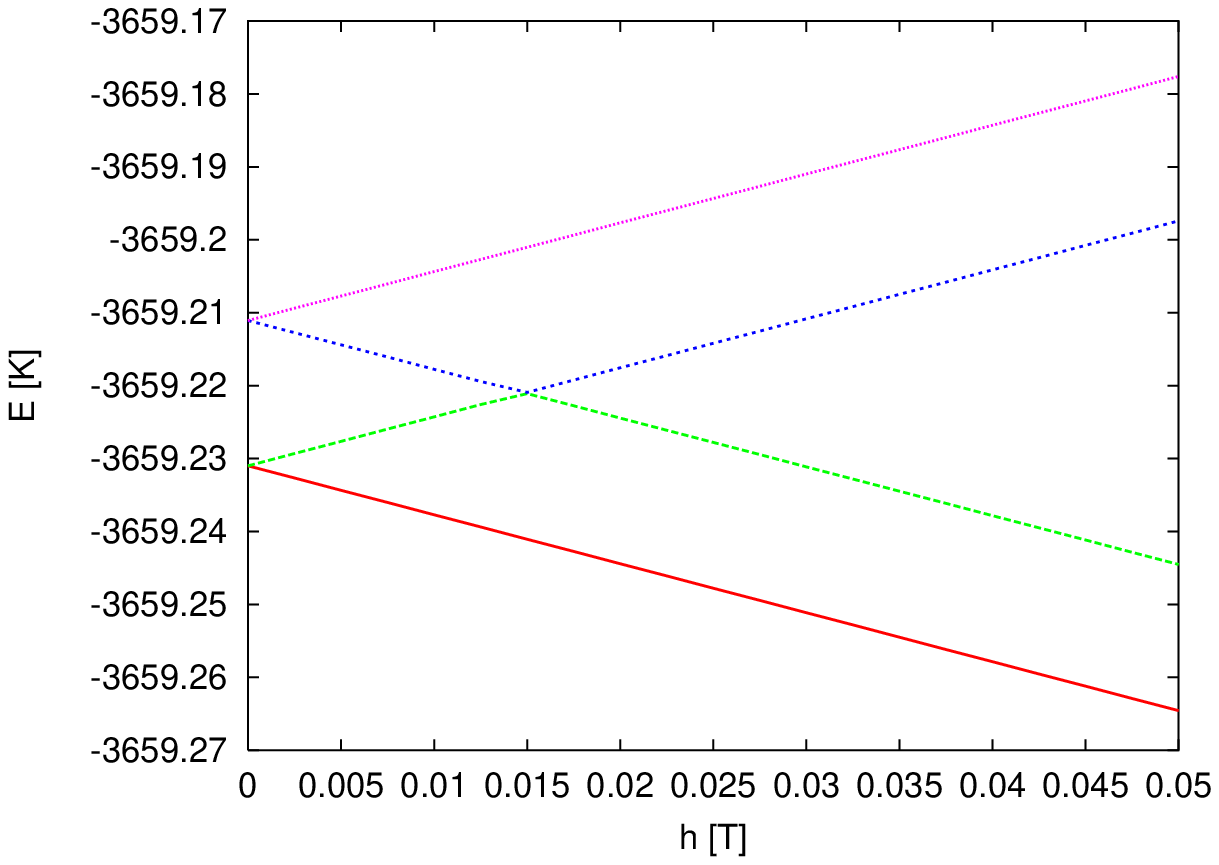}}
\end{picture}
\caption{%
Left:
The eight lowest energy levels of \V\ model (\ref{VHam})
with model parameters taken from Ref.~\cite{Konst}
as a function of the applied magnetic field ${\bf h}$
parallel to the $z$-axis.
The values of the DM vector are given in the text.
Right: Detailed view of the four lowest energy levels at $h\approx0$.}
\label{fig6}
\label{fig7}
\end{center}
\end{figure}

First we discuss our results for the lowest energy levels for
various sets of estimates for the model parameters
of the 15-spin model of \V~\cite{gat2,Rudra,Konst,Bouk0}.
For the model parameters given in Ref.~\cite{Rudra},
$J=-800$K, $J_1=J'=-54.4$K, and $J_2=J''=-160$K, $J_3=J_4=J_5=J_6=0$ and
in the absence of DM interactions,
we find for the energy gap between the ground state
and the first excited state at $h=0$ a value of 4.12478K. This reproduces
the result of Ref.~\cite{Rudra} and is also
in reasonable agreement with the experimental value
of approximately 3.7K~\cite{private}.

Following Ref.~\cite{RudraSeiji} we now take for the DM interaction parameters
$D^x_{1,2}=D^y_{1,2}=D^z_{1,2}=40$K,
which is approximately 5\% of the largest Heisenberg coupling.
Using the rotational symmetry of the hexagon it follows from Eq.(\ref{c3sym}) that
$D^x_{3,4}=14.641$K, $D^y_{3,4}=-54.641$K, $D^z_{3,4}=40$K and
$D^x_{5,6}=-54.641$K, $D^y_{5,6}=14.641$K, $D^z_{5,6}=40$K.
If the two hexagons are not equivalent we cannot use the
symmetry to reduce the number of free parameters (see below).
However, for simplicity, we may assume that the $(x,y)$-positions of the spins on the lower hexagons
differ from those on the upper hexagon by a rotation about $\pi/3$.
This yields for the remaining model parameters
$D^x_{10,11}=-14.641$K, $D^y_{10,11}=54.641$K, $D^z_{10,11}=40$K,
$D^x_{12,13}=-40$K, $D^y_{12,13}=-40$K, $D^z_{12,13}=40$K, and
$D^x_{14,15}=54.641$K, $D^y_{14,15}=-14.641$K, $D^z_{14,15}=40$K.
For this choice of model parameters,
the eight lowest energies of \V\ model (\ref{VHam})
for two values of the applied magnetic field ($h=0$ and $h=4T$) along the $z$-axis
can be found in Table~\ref{Vtab}.
From Table~\ref{Vtab} we see that for zero field, the DM interaction splits the
doubly-degenerate doublets of $S=1/2$ states into two doublets of $S=1/2$ states.
The difference in energy between the doubly-degenerate, first excited states and the
two-fold degenerate ground states has a value of $0.0085$K.
This value is much smaller than the experimental estimate of $0.05$K~\cite{Irinel4},
but of the same order of magnitude as the value cited in Ref.~\cite{Konst}.
The next four higher levels are $S=3/2$ states.
The energy-level splitting between the $S=3/2$ and $S=1/2$ states is
$4.1$K.
The value of this splitting weakly depends on the value of the DM interaction (results not shown).

\begin{table}[t]
\begin{center}
\caption{%
The eight lowest eigenvalues $E_i$ and total spin $S_i$ of the corresponding eigenstates of
the \V\ model (\ref{VHam})
with model parameters taken from Ref.~\cite{Rudra}
for two values of the external applied field ${\bf h}$ parallel to the $z$-axis.
The values of the DM vectors are given in the text.
The distance between $E_i$ and the exact eigenvalue closest to $E_i$ is
$\Delta_i=\langle\varphi_i|(H-E_i)^2|\varphi_i\rangle^{1/2}<\times10^{-9}$ for $i=1,\ldots,7$.}
\label{Vtab}
\begin{ruledtabular}\begin{tabular}{ccccc}
$i$ & $E_i(h=0)$ & $S_i(h=0)$ & $E_i(h=4T)$ & $S_i(h=4T)$ \\
\hline
  0 & -3679.53623744 &  0.51 &-3683.51181131& 1.50 \\
  1 & -3679.53623744 &  0.51 &-3682.21997451& 0.51 \\
  2 & -3679.52777009 &  0.51 &-3682.18488706& 0.53 \\
  3 & -3679.52777009 &  0.51 &-3678.11784886& 1.50 \\
  4 & -3675.42943612 &  1.50 &-3676.84225573& 0.52 \\
  5 & -3675.42943612 &  1.50 &-3676.83951808& 0.51 \\
  6 & -3675.42325141 &  1.50 &-3672.74011178& 1.50 \\
  7 & -3675.42325141 &  1.50 &-3667.37940477& 1.50 \\
 \end{tabular}
 \end{ruledtabular}
 \end{center}
 \end{table}

Following Ref.~\cite{Konst}, we take
$J=-800$K, $J_1=J'=-225$K, $J_2=J''=-350$K, and $J_3=J_4=J_5=J_6=0$.
In the absence of DM interactions,
we find that the energy gap between the four-fold degenerate ground state
and the first excited state is 3.61K, in full agreement with the result of Ref.~\cite{Konst}.
Note that this value of the gap is fairly close to the experimental value
of 3.7K~\cite{private}.
Taking for the DM interactions
$D^x_{1,2}=D^x_{14,15}=25$K,
$D^x_{3,4}=D^x_{5,6}=D^x_{10,11}=D^x_{12,13}=-12.5$K,
$D^y_{3,4}=-D^y_{5,6}=-D^y_{10,11}=D^y_{12,13}=-21.5$K,
our calculation for the splitting between the two doubly-degenerate
$S=1/2$ levels yields $0.0037$K, about a factor of two
larger than the value cited in Ref.~\cite{Konst}.
For the energy splitting between the $S=1/2$ and $S=3/2$ levels
we obtain $3.616$K instead of the value $3.618$K given in Ref.~\cite{Konst}.
These differences seem to suggest that a perturbation approach~\cite{Konst} for the
DM interaction may not be sufficiently accurate for quantitative purposes~\cite{kosty}.

The most advanced estimation of the model parameters is given in Ref.~\cite{Bouk0}.
Taking $J=-809$K, $J'=-120$K, $J''=120$K, $J_1=-30$K, $J_2=-122$K,
$J_3=-3$K, $J_4=-11$K, $J_5=-3$K, $J_6=-2$K (see Table I in Ref.~\cite{Bouk0})
yields an energy gap of 4.915K, in agreement with Ref.~\cite{Bouk0}.
Adding the DM interaction with parameters
$D^x_{1,2}=D^x_{14,15}=25$K,
$D^x_{3,4}=D^x_{5,6}=D^x_{10,11}=D^x_{12,13}=-12.5$K,
$D^y_{3,4}=-D^y_{5,6}=-D^y_{10,11}=D^y_{12,13}=-21.5$K,
a level repulsion appears at $h\approx3.6$T where the $S=1/2$ and $S=3/2$ states mix
(results not shown).
Although the qualitative features of the energy-level diagram for this
set of model parameters also agree with what one would expect on the basis of experiments,
the field at which the states $|1/2,1/2\rangle$ and $|3/2,3/2\rangle$ cross, $h\approx3.6$T,
does not compare well to the experimental estimate $h\approx2.8$T.

In Fig.~\ref{fig6} we show the results
for $J=-800$K, $J_1=J'=-225$K, and $J_2=J''=-350$K~\cite{Konst} and
the same DM interaction parameters as those used to obtain the results of Table ~\ref{Vtab}.
For the energy gap at zero field,
we find 3.7K (instead of 4.1K for the $J$'s of Ref.~\cite{Rudra}), in good agreement
with the experimental estimate of 3.7K~\cite{private}.
The transition between the states
$|1/2,1/2\rangle$ and $|3/2,3/2\rangle$ takes place at
$h\approx2.8$T (instead of $h\approx3.0$T for the $J$'s of Ref.~\cite{Rudra}),
also in good agreement with the experimental value $2.8$T.

The results for the zero-field energy gap suggest that there are many different sets
of model parameters
that approximately reproduce the experimental gap between the singlet and triplet states.
However, as we show below, the energy gap at zero field not necessarily corresponds
to the gap of a level repulsion that is required for the adiabatic change of the
magnetization.

\begin{figure}[t]
\begin{center}
\setlength{\unitlength}{1cm}
\begin{picture}(14,6)
\put(-2.,0.){\includegraphics[width=9cm]{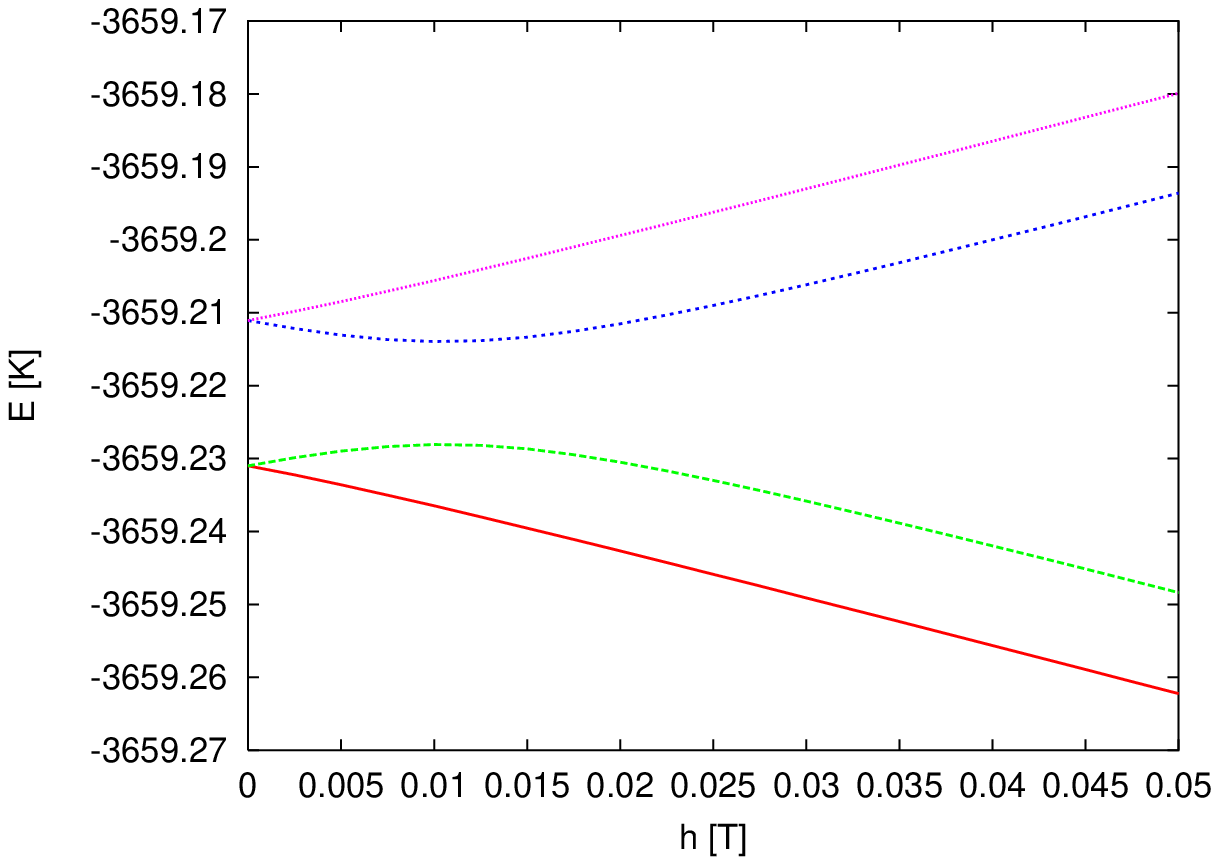}}
\put(7.,0.){\includegraphics[width=9cm]{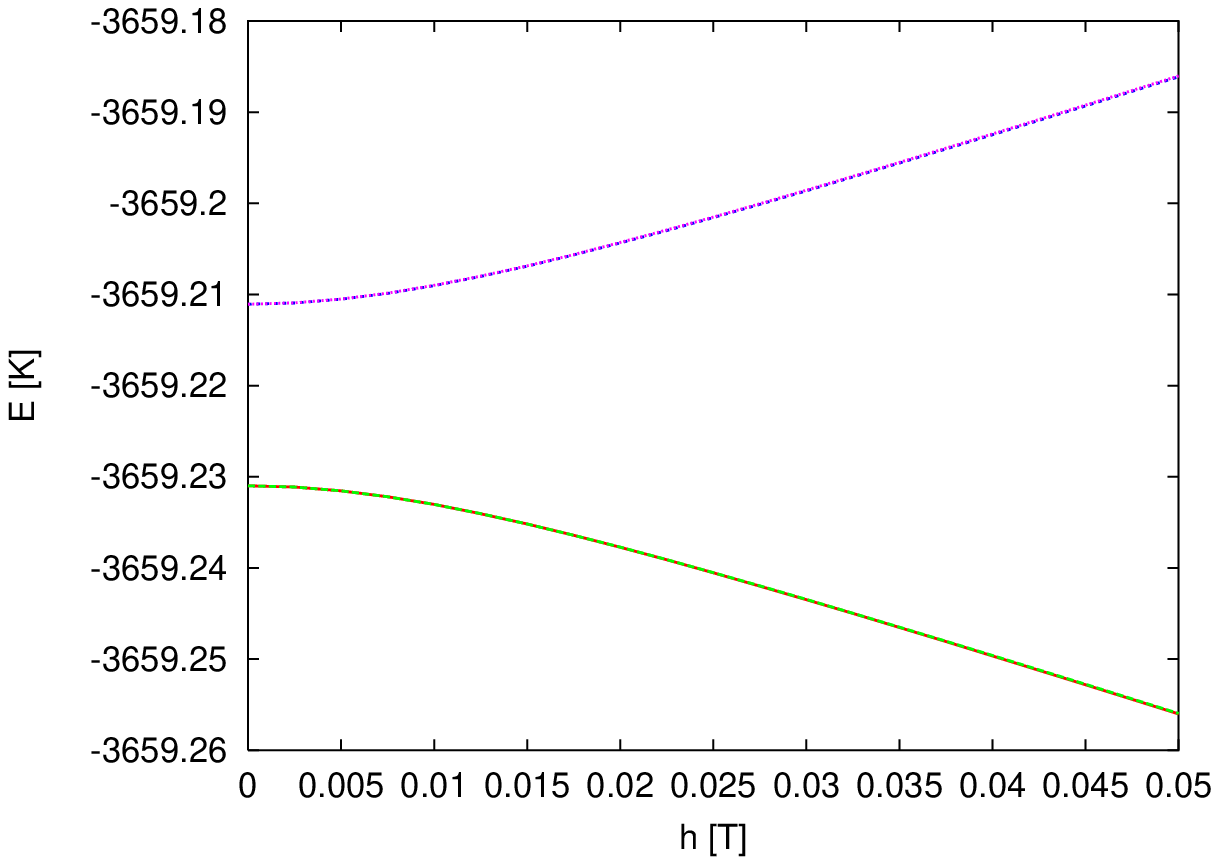}}
\end{picture}
\caption{%
Same as the right panel in Fig.~\ref{fig6} except that
the applied magnetic field ${\bf h}=h(1,0,1)/\sqrt{2}$ is tilted by 45$^\circ$ with respect to the $z$-axis (left)
and ${\bf h}$ is along the $x$-axis (right).}
\label{fig8}
\label{fig9}
\end{center}
\end{figure}

We now focus on the relation between the direction of the
magnetic field and the structure of the energy levels.
As shown in the right panel of Fig.~\ref{fig6},
when the field is parallel to the symmetry axis
the energy levels of $M=\pm 1/2$ simply cross, just as in the case of the left panel of Fig.~\ref{fig1}.
In Fig.~\ref{fig8} we present results for the cases where the angle between the applied field
and the $z$-axis is 45$^{\circ}$ and 90$^{\circ}$, respectively.
Clearly, we find the same type of dependence of the energy levels on the
angle of the field as in the case of the three-spin model
(see right panel of Fig.~\ref{fig3} and right panel of Fig.~\ref{fig1}, respectively).
Exactly the same qualitative features are obtained
for the other sets of parameters discussed above (results not shown).

\begin{figure}[t]
\begin{center}
\setlength{\unitlength}{1cm}
\begin{picture}(14,6)
\put(-2.,0.){\includegraphics[width=9cm]{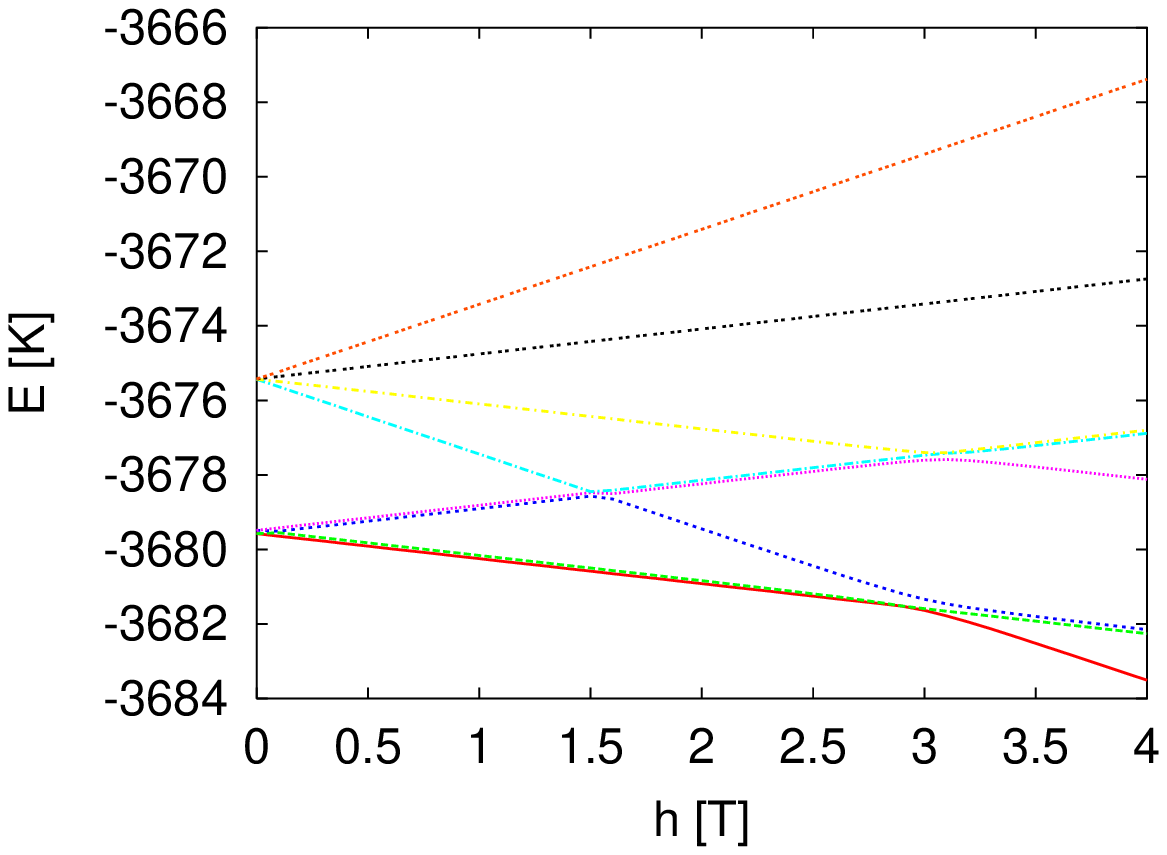}}
\put(7.,0.){\includegraphics[width=9cm]{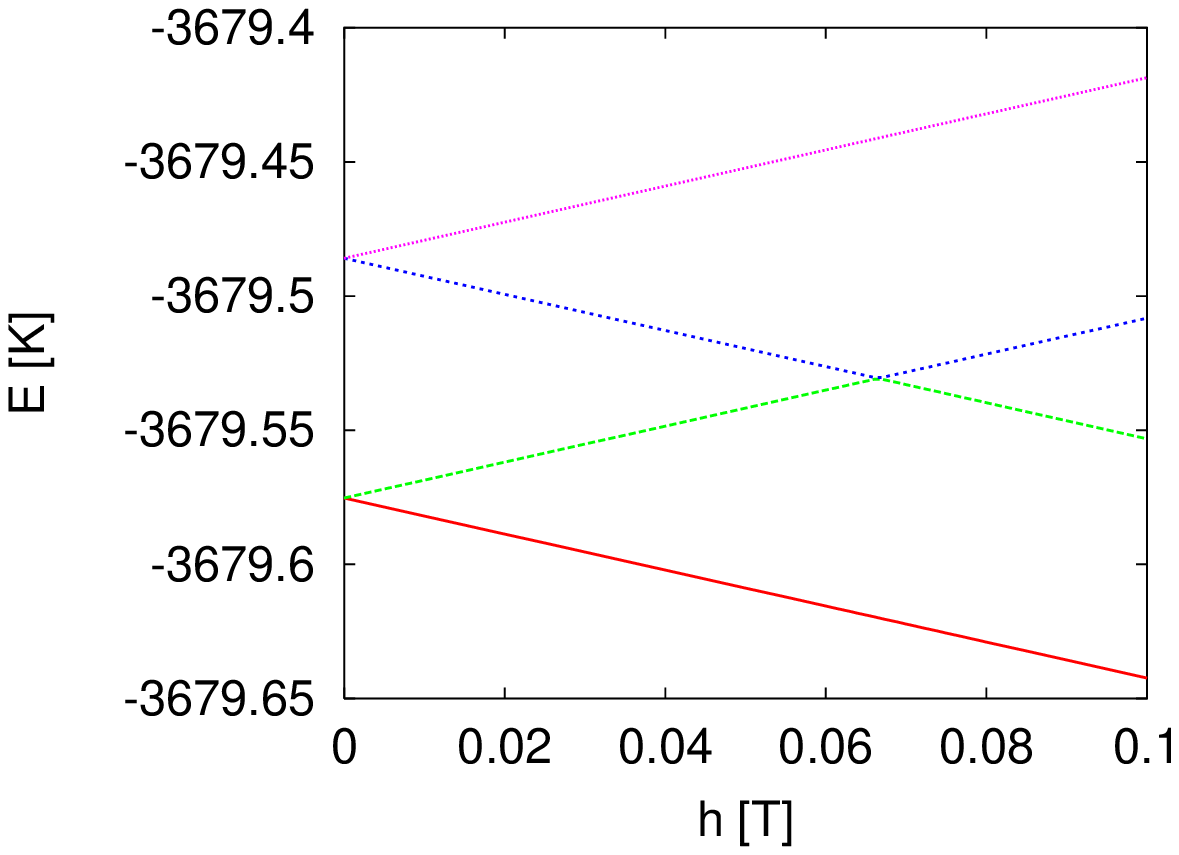}}
\end{picture}
\caption{%
Left:
The eight lowest energy levels of \V\ model (\ref{VHam})
with model parameters taken from Ref.~\cite{Rudra}
assuming that the DM vectors of the upper and lower hexagon
are related to each other by a 180$^\circ$ rotation around
a vector that passes through V-atom number 7 (see Fig.~\ref{fig5}) of the triangle
and the middle of the line connecting the two other V-atoms
of the triangle.
The applied magnetic field ${\bf h}$ is parallel to the $z$-axis.
Right: Detailed view of the $h$-dependence of the four lowest energy levels.}
\label{fig12}
\label{fig13}
\end{center}
\end{figure}

\begin{figure}[t]
\begin{center}
\setlength{\unitlength}{1cm}
\begin{picture}(14,6)
\put(-2.,0.){\includegraphics[width=9cm]{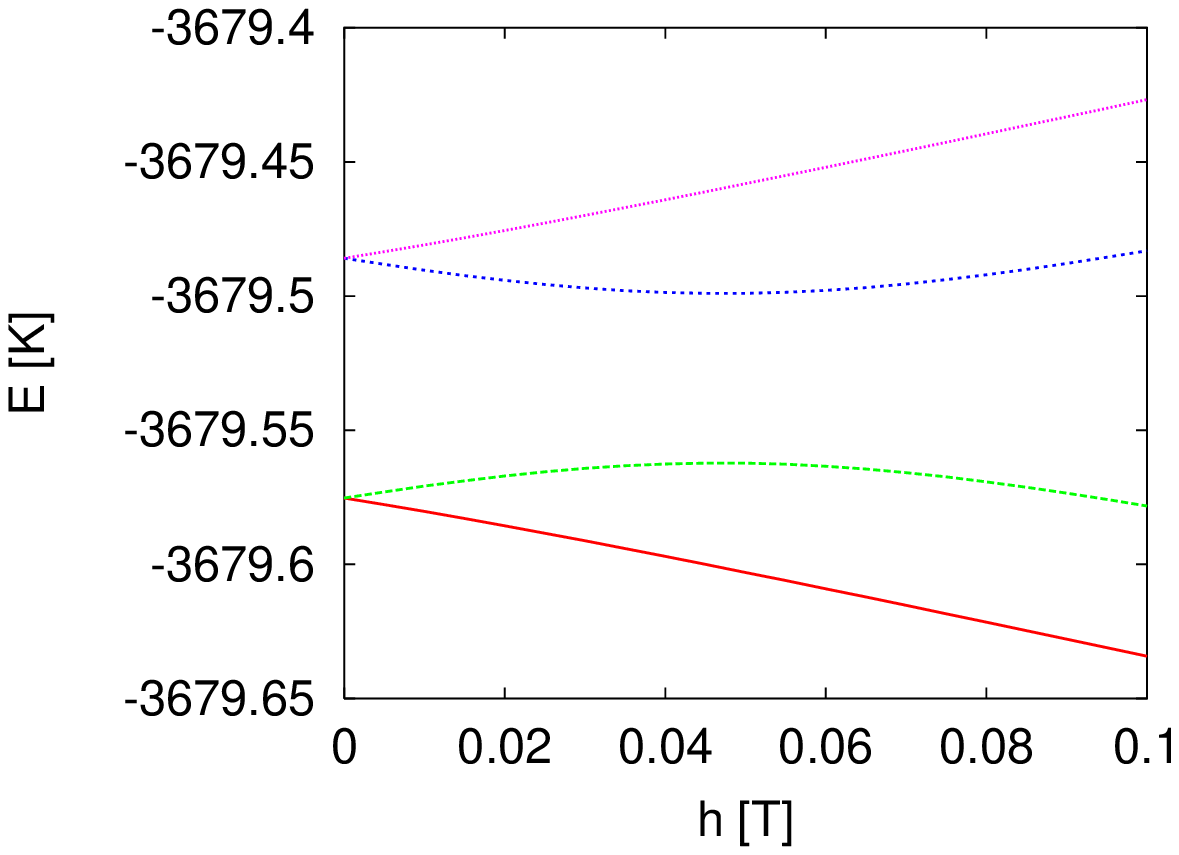}}
\put(7.,0.){\includegraphics[width=9cm]{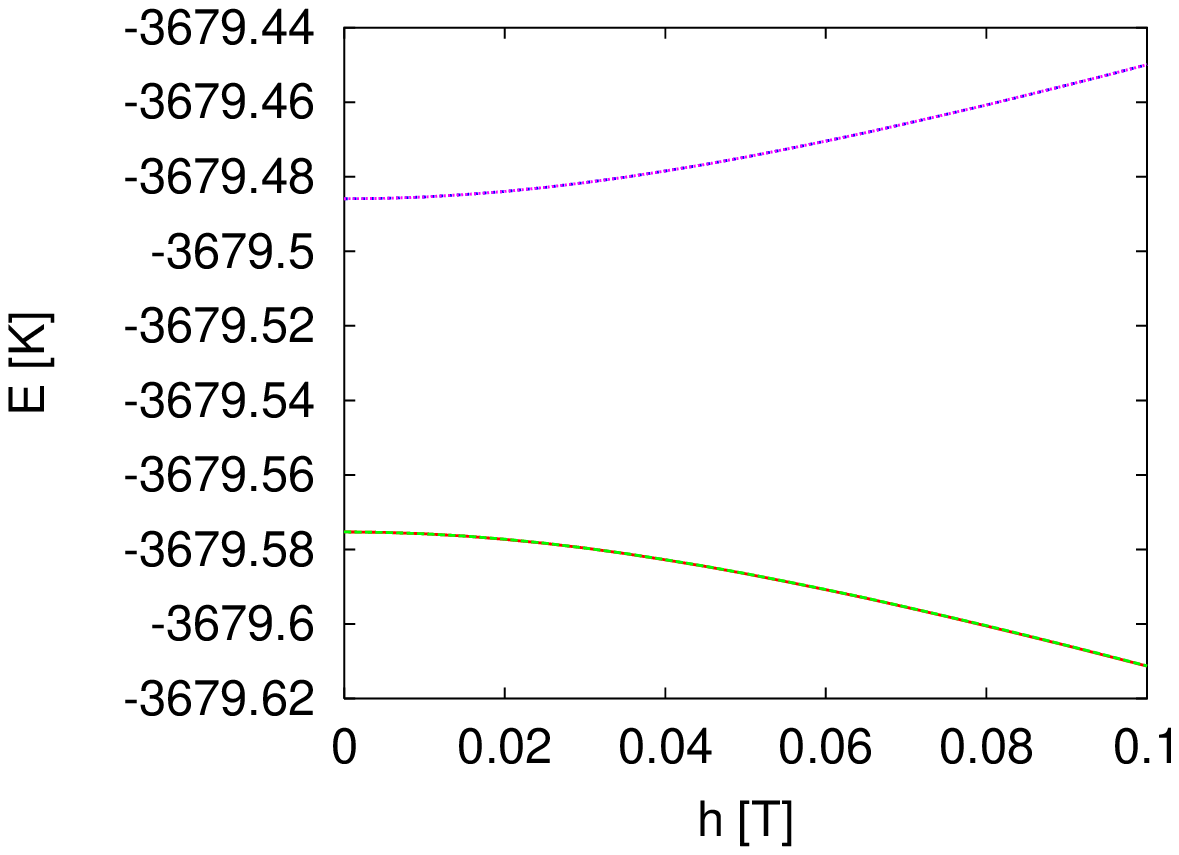}}
\end{picture}
\caption{%
Same as the right panel in Fig.~\ref{fig12} except that
the applied magnetic field ${\bf h}=h(1,0,1)/\sqrt{2}$ is tilted by 45$^\circ$ with respect to the $z$-axis (left)
and ${\bf h}$ is along the $x$-axis (right).}
\label{fig14}
\label{fig15}
\end{center}
\end{figure}

Up to now,  we used DM vectors that satisfy the rotational symmetry of a hexagon, and we
took the same DM vectors for the other hexagon for simplicity~\cite{Konst}.
However, if there is some symmetry which connects the upper and lower hexagon,
we have a relation between the DM vectors on both hexagons.
In concert with the relations between the exchange couplings,
let us assume that the upper and lower hexagon are related to each other by a
180$^\circ$ rotation around a vector that passes through V-atom number 7 (see Fig.~\ref{fig5})
and the middle of the line connecting the two other V-atoms of the triangle.
This symmetry operation $X$
transforms the sites \{(1,2), (3,4), (5,6)\} into \{(14,15), (10,11), (12,13)\}.
If we place the $y$-axis along the line through V-atom number 7 and through the middle of the
line connecting V-atoms 8 and 9, and if we take
${\bf D}_{1,2}=(D_x,D_y,D_z)$ as the reference DM vector,
the other DM vectors are given by
${\bf D}_{3,4}=R^2{\bf D}_{1,2}$, ${\bf D}_{5,6}=R{\bf D}_{1,2}$,
${\bf D}_{14,15}=X{\bf D}_{1,2}$, ${\bf D}_{10,11}=RX{\bf D}_{1,2}$,
and ${\bf D}_{12,13}=R^2X{\bf D}_{1,2}$. Here $R$ denotes a rotation of the hexagons
around $2\pi/3$ in the plane of the hexagons.
In Figs.~\ref{fig12} and \ref{fig15} we show the energy level
diagrams for the set of parameters taken from Ref.~\cite{Rudra},
with the additional constraint imposed by the symmetry.
Qualitatively the results are similar to those
obtained in the absence of a symmetry that relates the upper and lower hexagons
(see Figs.~\ref{fig6} and \ref{fig8}).

Summarizing: As in the case of the three-spin model, all our results for the 15-spin \V\ model
clearly demonstrate that the mixing of levels strongly depends on
the direction of the magnetic field.
It seems therefore that this dependence is a generic feature of the DM interaction.

\section{Effects of lower symmetry}
\label{sec4}

\begin{figure}[t]
\begin{center}
\includegraphics[width=9cm]{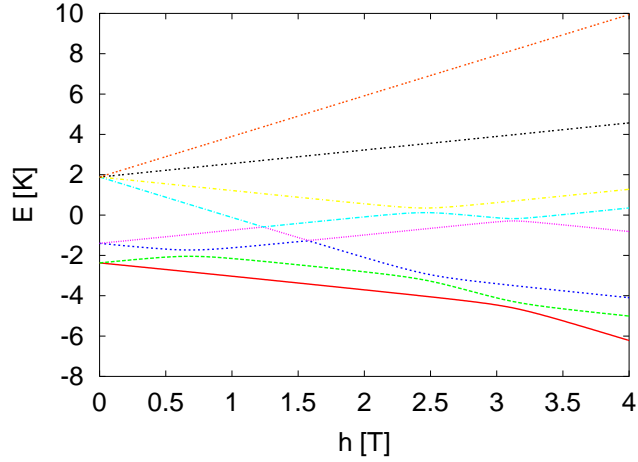}
\caption{%
Energy level diagram of the Hamiltonian (\ref{HD})
of a distorted triangle with model parameters
$J_{1,2}=-2.5$K, $J_{2,3}=-2.0$K, $J_{3,1}=-3.0$K, and $D_x=D_y=D_z=25$K.
The applied magnetic field ${\bf h}=h(1,0,1)/\sqrt{2}$ is tilted by 45$^\circ$ with respect to the $z$-axis.
}
\label{LOWSYM}
\end{center}
\end{figure}

\begin{figure}[t]
\begin{center}
\setlength{\unitlength}{1cm}
\begin{picture}(14,6)
\put(-2.,0.){\includegraphics[width=9cm]{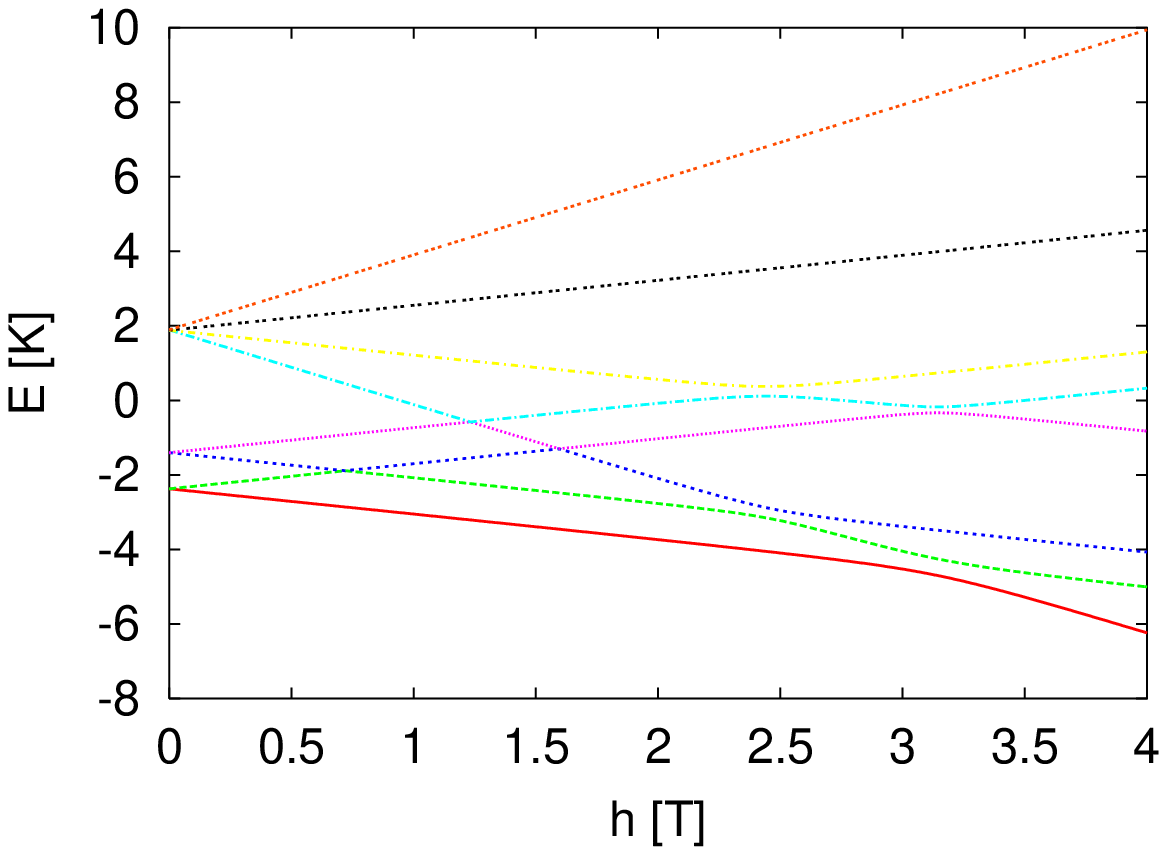}}
\put(7.,0.){\includegraphics[width=9cm]{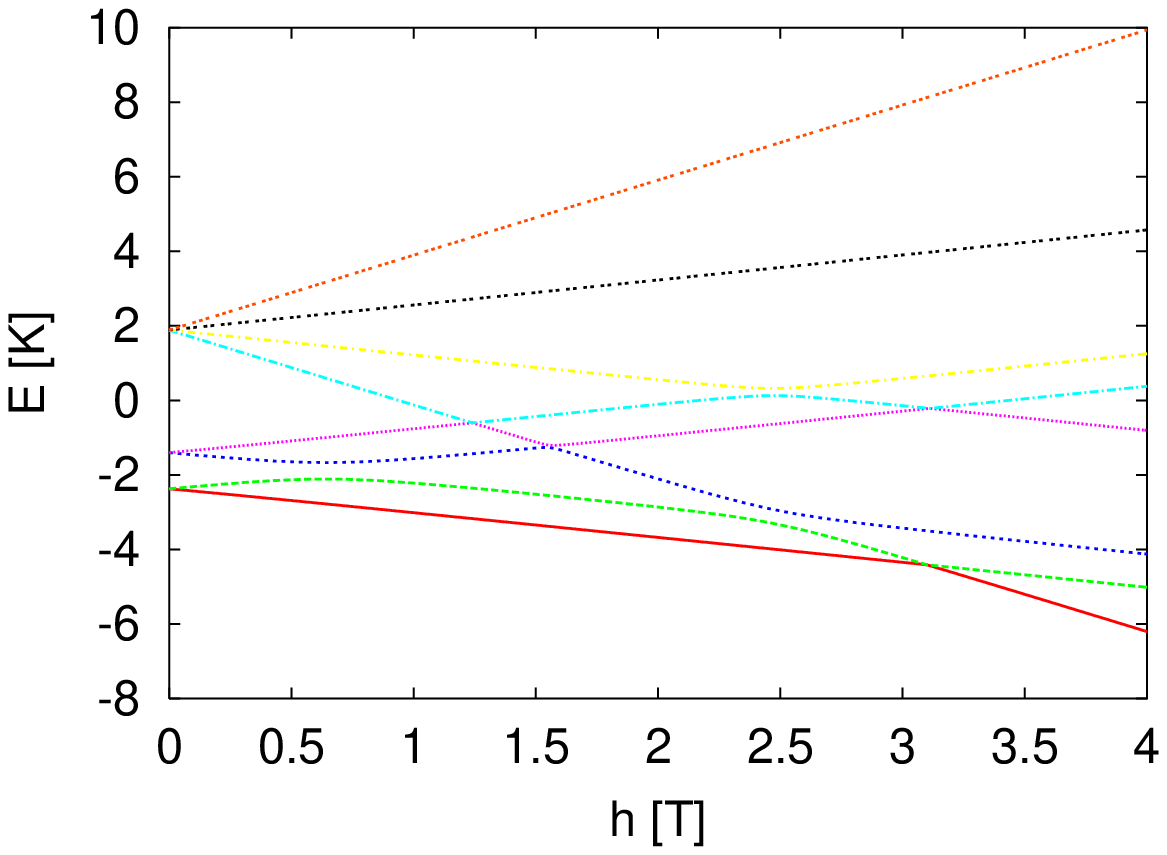}}
\end{picture}
\caption{%
Same as Fig.~\ref{LOWSYM} except that
the applied magnetic field ${\bf h}$ is parallel to the $z$-axis (left) and
along the $x$-axis (right).}
\label{LOWSYM0}
\label{LOWSYM90}
\end{center}
\end{figure}

We now discuss the effects of distortion of the triangle and anisotropic exchange interactions in the
triangle-model of \V\ on the energy level diagram for $h\approx 0$.
When the triangle is distorted ($J_{1,2}\neq J_{2,3}\neq J_{3,1}\neq J_{1,2}$)
the degeneracy of the two doublets at $h=0$ is lifted, even if $D_x=D_y=D_z=0$.
In Figs.~\ref{LOWSYM} and ~\ref{LOWSYM90} we show data for
$J_{1,2}=-2.5$K, $J_{2,3}=-2.0$K and $J_{3,1}=-3.0$K and $D_x=D_y=D_z=0.25$K
for the applied magnetic field ${\bf h}=h(1,0,1)/\sqrt{2}$ tilted by 45$^\circ$ with respect to the $z$-axis,
and for field directions parallel to the $z$-axis and along the $x$-axis.
Unless the field is parallel to the $z$-axis, the energy level diagrams are
qualitatively similar to the ones of the undistorted triangle
(see Figs.~\ref{fig1} and ~\ref{fig3}),
that is, crossings at $h=0$ and level repulsions close to $h=0$ but not at $h=0$.
However, in the case of an undistorted triangle with anisotropic, antiferromagnetic exchange interactions
(i.e. different for the $x$, $y$, and $z$-components of the spins) and $D_x=D_y=D_z=0.25$K,
the energy level diagrams are qualitatively similar to the ones of the undistorted triangle (results not shown).

\begin{figure}[t]
\begin{center}
\setlength{\unitlength}{1cm}
\begin{picture}(14,8)
\put(-1.5,0.){\includegraphics[width=7cm]{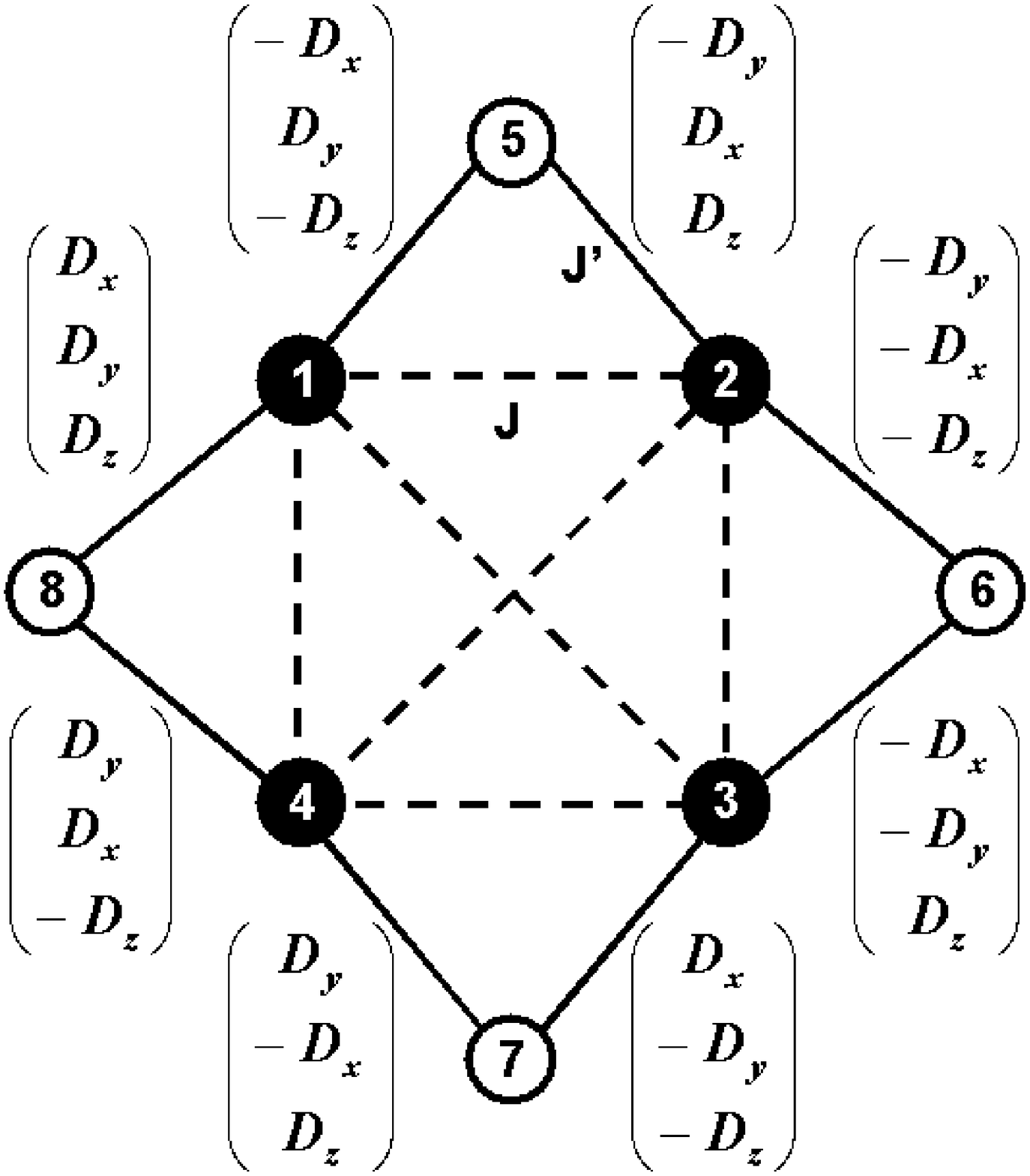}}
\put(7.,0.){\includegraphics[width=9cm]{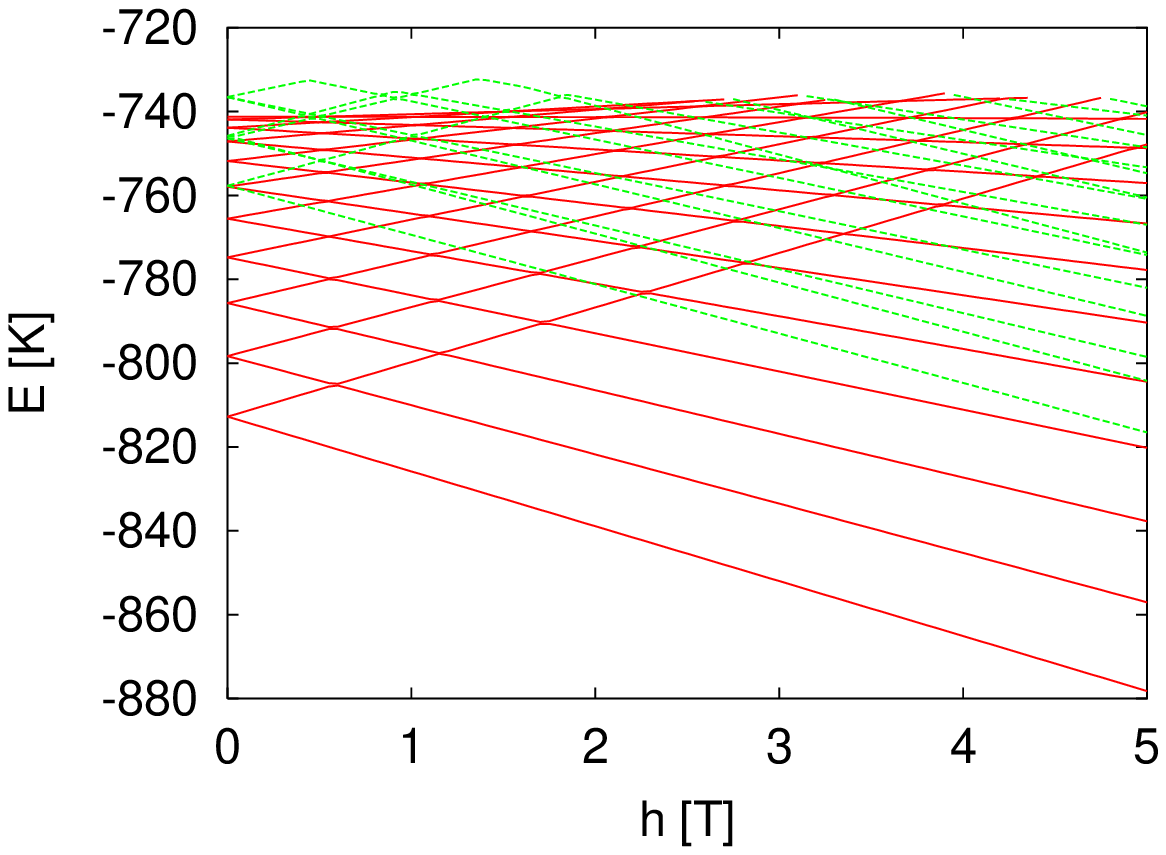}}
\end{picture}
\caption{%
Left:
Schematic diagram of the magnetic interactions of the simplified
model (\ref{MnHam}) of the \Mn\ molecule~\cite{Misha1}.
Black circles: $S=1/2$; open circles: $S=2$.
Also shown are the DM vectors (for $i<j$ and ${\bf D}_{i,j}=-{\bf D}_{j,i}$).
Right:
The 21 lowest energy levels of the \Mn\ model (\ref{MnHam}) as a function of the applied magnetic field
${\bf h}$.
Solid lines: Eigenstates with $S\approx10$;
dashed lines: Eigenstates with $S\approx9$.
The applied magnetic field ${\bf h}$ is parallel to the $z$-axis.
}
\label{fig17}
\label{fig18}
\end{center}
\end{figure}

\section{Manganese complex: $\hbox{Mn}_{12}$}\label{sec5}

The four inner Mn$^{+4}$ ions in the \Mn\ molecule
(Mn$_{12}$(CH$_{3}$COO)$_{16}$(H$_{2}$O)$_{4}$O$_{12}\cdot$2CH$_{3}$COOH$\cdot$4H$_{2}$O),
have spin $S=3/2$. The other eight Mn$^{+3}$ ions have spin $S=2$.
The dimension of the Hilbert space of this system is $4^4\times5^8=10^{8}$.
If the total magnetization is a conserved quantity, it can be used to
block-diagonalize the Hamiltonian, allowing a numerical study of models of this size~\cite{Raghu,Regnault}.
However, to study the adiabatic change of magnetization, we have to take into account
all the states, and the dimension of the matrix becomes prohibitively large.
Thus we need to simplify the model in order to reduce the dimension.
A drastic reduction of the number of spin states can be achieved by assuming that the strong
antiferromagnetic Heisenberg interaction $J^{\prime}$ between an inner ion and its outer neighbor
allows the replacement of the magnetic moment of an inner ion by an effective $S=1/2$ moment.
The schematic diagram of this simplified (but still complicated) model~\cite{Misha1} is shown in
the left panel of Fig.~\ref{fig17}.
The dimension of the Hilbert space of this model is $2^4\times5^4=10^{4}$.
In the following we study this simplified model.

The Hamiltonian for the magnetic interactions of the simplified \Mn\ model can be written as~\cite{Misha1}

\begin{eqnarray}
{\cal H}&=& -J\Bigl(\sum_{i=1}^4 {\bf S}_i \Bigr)^2
  -J'\sum_{(i,j)\in\, {\cal B}} {\bf S}_i \cdot{\bf S}_j
  -K_z\sum_{i=1}^4 \left(S_{i+4}^z\right)^2 
  + \sum_{(i,j)\in\, {\cal B}} {\bf D}_{i,j} \cdot
     [{\bf S}_i\times {\bf S}_j]
     - \sum_{i=1}^{8} {\bf h}\cdot{\bf S}_{i},
\label{MnHam}
\end{eqnarray}
where the index $1\le i\le 4$ ($5\le i\le 8$) refers to $S=1/2$ ($S=2$) spins and
${\cal B}$ denotes the set of pairs
${\cal B}=\left\{(1,5),(1,8),(2,5),(2,6),(3,6),(3,7),(4,6),(4,8)\right\}$.
The first two terms describe the isotropic Heisenberg exchange between the spins.
The third term ($K_z$) describes the single-ion easy-axis anisotropy of $S=2$ spins.
The fourth term represents the anti-symmetric DM interaction in \Mn.
The vector ${\bf D}_{i,j}$ determines the DM interaction between the $i$-th $S=1/2$ spin and the $j$-th $S=2$ spin.
The last term describes the interaction of the spins with the external magnetic
field ${\bf h}$.
Note that the factor $g\mu_B$ is absorbed in our definition of ${\bf h}$.
Model (\ref{MnHam}) reproduces experimental data,
such as the splitting of the neutron scattering peaks, results of EPR measurements and the
temperature dependence of the magnetic susceptibility~\cite{Misha1}.

The first three terms in Hamiltonian (\ref{MnHam}) conserve the
$z$-component of the total spin $M^z=\sum_{i=1}^{8} S^z_{i}$.
The DM interaction, on the other hand, mixes states with different total spin.
Hence, the DM interaction can change level crossings into level repulsions.
Therefore, the presence of the DM interaction may be sufficient
to explain the experimentally observed adiabatic changes of the magnetization.

The four-fold rotational-reflection symmetry ($S_4$)
of the \Mn\ molecule imposes some relations between the DM-vectors.
There are only three independent DM-parameters~\cite{Misha1}:
$D_x \equiv D^x_{1,8}$, $D_y\equiv D^y_{1,8}$, and $D_z \equiv D^z_{1,8}$,
as indicated in Fig.~\ref{fig17}.
The parameters of model (\ref{MnHam}) have been estimated by comparing experimental and theoretical data.
In this paper we will use the parameter set B from Ref.~\cite{Misha1,Hans1}:
$J=23.8$K, $J'=79.2$K, $K_z=5.72$K, $D_x=22$K, $D_y=0$, and $D_z=10$K.
Although the amount of available data is not sufficient to fix all these parameters accurately,
we expect that the general trends in the energy-level diagram will not change drastically
if these parameters change relatively little.

In Table~\ref{Mntab} we present results for the energy and total spin
of the 21 lowest states for $h=0$ and $h=5$T
(in these calculations ${\bf h}$ is parallel to the $z$-axis).
The numerical results obtained by full exact diagonalization (LAPACK), the
Lanczos method with full orthogonalization (LFO) (see appendix B)
and the Chebyshev-polynomial-based projector method (CPP) (see appendix B) are
the same to working precision (about 13 digits).
Clearly there are states with total spin 8, 9 and 10 within these 21 lowest eigenstates.
Although the total magnetization is not a good quantum number, we can
label the various eigenstates by their (calculated) magnetization.

The $S=10$ single-spin model for \Mn\

\begin{eqnarray}
{\cal H}&=& -D (S^z)^2-h S^z,
\label{SSHam}
\end{eqnarray}
where $D$ denotes the uniaxial anisotropy,
is often used as a starting point to interpret experimental
results~\cite{Friedman,Thomas,Perenboom,Irinel3,Pohjola,Rudra}.
The energy levels of this model exhibit crossings at the resonant fields
$h=\pm Dn$ for $n=-10,\ldots,10$,
in qualitative agreement with our numerical results (shown in the right panel of Fig.~\ref{fig18})
for the microscopic model (\ref{MnHam}).
A fit of the first eight level crossings
of model (\ref{SSHam}) to the data of Fig.~\ref{fig18} yields
$D\approx0.74$K, in good agreement with experiments~\cite{Friedman,Thomas}.
The Hamiltonian of the single-spin model (\ref{SSHam}) commutes with the magnetization $S^z$ and
therefore its energy diagram only displays level crossings, no level repulsions.
Adding anisotropy of the form $C(S_+^4 + S_-^4)$ to model (\ref{SSHam}) changes
the estimated value of $D$ and leads to level repulsions when the magnetization changes by 4
~\cite{Misha1,BARR97,LUIS98,HILL98,BARC02}.

It is also of interest to compare the level splitting at $h=0$
obtained by lowest-order degenerate perturbation theory
of model (\ref{SSHam}) with fourth-order anisotropy of the form $S_+^4 + S_-^4$~\cite{GARA91,GARA97,LUIS98}

\begin{eqnarray}
\Delta E_{l+1,l}=32D\left(\frac{C}{16D}\right)^{m/2}\frac{1}{((m/2-1)!)^2}\frac{(S+m)!}{(S-m)!},
\label{Luisdelta}
\end{eqnarray}
for $m$ even ($\Delta E_{l+1,l}=0$ for $m$ odd)
with the result of the microscopic model calculation based on model (\ref{MnHam}).
In Eq.~(\ref{Luisdelta})
$l$ denotes the perturbed eigenstates in increasing order of energy and $m$ is the magnetic quantum number
of the unperturbed states~\cite{LUIS98}.
Using the values $D\approx0.69$K, $C/D\approx5.7\times10^{-5}$
obtained by fitting the single-spin model to experimental data~\cite{LUIS98},
the energy gap for $m=6$ is given by

\begin{eqnarray}
\Delta E_{13,12}\approx0.00022K.
\end{eqnarray}
Taking into account that (because of the presence of $S\approx 9$ states)
the corresponding $S\approx10$ levels of model (\ref{MnHam})
are the 14-th and 15-th lowest energy level,
Table~\ref{Mntab} shows that for $h=0$, $\Delta E_{13,12}=0.00042$K.
In view of the uncertainties on the estimates of the various model parameters,
the difference of only a factor of two is remarkably small.
From this comparison, we may conclude that the DM interaction leads to energy gaps that are of the same order
of magnitude as the gaps due to the fourth-order terms $S_+^4 + S_-^4$ in the single-spin model.

\begin{table}[t]
\begin{center}
\caption{%
The 21 lowest eigenvalues $E_i$ and total spin $S_i$ of the corresponding eigenstates of
the \Mn\ model (\ref{MnHam}) for two values of the external applied field ${\bf h}$ along the $z$-axis.
The distance between $E_i$ and the exact eigenvalue closest to $E_i$ is
$\Delta_i=\langle\varphi_i|(H-E_i)^2|\varphi_i\rangle^{1/2}<10^{-10}$ for $i=1,\ldots,7$.
Note that for $h=0$ the levels 12,13,18,19, and 20 belong to the $S\approx9$ subspace
and not to the $S\approx10$ subspace.}
\label{Mntab}
\begin{ruledtabular}\begin{tabular}{ccccc}
$i$ & $E_i(h=0)$ & $S_i(h=0)$ & $E_i(h=5T)$ & $S_i(h=5T)$ \\
\hline
  0 &  -812.771882673675 & 9.72 & -878.203468556749  & 9.77\\
  1 &  -812.771882673460 & 9.72 & -857.042137859145  & 9.78\\
  2 &  -798.326618260922 & 9.72 & -837.727273846391  & 9.76\\
  3 &  -798.326618261122 & 9.72 & -820.218084451590  & 9.73\\
  4 &  -785.677644659194 & 9.70 & -816.530388063144  & 8.82\\
  5 &  -785.677644658983 & 9.70 & -804.449056631056  & 9.69\\
  6 &  -774.774953284432 & 9.68 & -804.242954124067  & 8.77\\
  7 &  -774.774953284294 & 9.68 & -798.519706376890  & 8.82\\
  8 &  -765.549187817101 & 9.65 & -790.336285910398  & 9.65\\
  9 &  -765.549187333902 & 9.65 & -788.717912654407  & 8.78\\
 10 &  -757.919915510036 & 9.61 & -782.037906282298  & 8.82\\
 11 &  -757.919915509970 & 9.61 & -777.789121186874  & 9.60\\
 12 &  -757.673722613912 & 8.77 & -774.222534072884  & 8.79\\
 13 &  -757.673722613981 & 8.77 & -773.614730624955  & 8.80\\
 14 &  -751.806498916496 & 9.57 & -766.993852023410  & 8.80\\
 15 &  -751.806072514140 & 9.57 & -766.720405060281  & 9.55\\
 16 &  -747.135398548595 & 9.54 & -760.807893152451  & 8.79\\
 17 &  -747.135398548602 & 9.54 & -760.482768227423  & 8.12\\
 18 &  -746.357522623039 & 8.77 & -757.060762637193  & 9.51\\
 19 &  -746.357522623082 & 8.77 & -754.700878489864  & 8.59\\
 20 &  -745.778951523327 & 8.70 & -753.310517350023  & 8.78\\
 \end{tabular}
 \end{ruledtabular}
 \end{center}
 \end{table}

In the right panel of Fig.~\ref{fig18} we show the results for the lowest 21 energy levels of the
\Mn\ model as a function of the applied magnetic field as obtained by LFO.
The applied magnetic field is parallel to the $z$-axis.
In Fig.~\ref{fig18} solid (dashed) lines represent eigenstates with $S\approx10$ (9) (within an error of about 10\%).
Also eigenstates with $S\approx8$ appear for $h>4$ but these
are not shown for clarity.
For the \Mn\ model, the DM induced energy splittings
between the $S\approx 10$, $M\approx-10$ state and other states are less than 10$^{-6}$K.
Adding an extra transverse field by tilting the ${\bf h}$-field by 5$^\circ$
with respect to the $z$-axis does not change this conclusion (results not shown).

Finally, we added to model (\ref{MnHam}) the next-to-lowest order relativistic correction to
the local anisotropy that is compatible with the symmetry of the square~\cite{Misha1}

\begin{eqnarray}
{\cal H}_1&=& K_1[(S^x_1)^2+(S^y_2)^2+(S^x_3)^2+(S^y_4)^2].
\label{K1}
\end{eqnarray}
Although we took a perhaps unrealistically large value of $K_1$ ($K_1=K_z/2$),
we were unable to detect energy level repulsions
up to the $M\approx-10$, $M\approx3$ transition (results not shown).
On the other hand in experiments~\cite{Friedman,Thomas,Perenboom},
adiabatic changes of the magnetization have been observed
at $h\approx3.4$T
($M_z\approx-10\rightarrow M_z\approx4$) and $h\approx3.9$T
($M_z\approx-10\rightarrow M_z\approx3$) and the magnitude of the energy splittings is
of the order of 10 nK~\cite{private}.
The precision of the present calculations is about $10^{-6}$K.
Thus, it is consistent that within the (very high) resolution in the ${\bf h}$-field and
the 13-digit precision of the calculation, no information about the gap
could be extracted.
The algorithms developed for the work presented in this paper can
be used for 33-digit calculations without modification
and we leave the calculation of the splittings for future work.

\section{Discussion}\label{sec6}

We have studied the dependence of the energy level diagrams, with level repulsions
due to the Dzyaloshinskii-Moriya interaction, on the direction of the applied magnetic field.
We found that the dependence on the direction of the magnetic field
seems to be generic, at least if the system has C$_3$ symmetry.
Our numerical data
suggest that the three-spin model reproduces the main features of the 15-spin model of \V.
The presence of the Dzyaloshinskii-Moriya interaction
allows for adiabatic changes of the magnetization but,
according to our calculations, the value of the resonant field
for the $|1/2,-1/2\rangle$ to $|1/2,1/2\rangle$ transition
changes with the direction of the magnetic field.
The Dzyaloshinskii-Moriya interaction not only lifts the degeneracy but,
depending on the direction of the field with respect to the symmetry axis,
also shifts the resonant point away from $h=0$.

The butterfly hysteresis loop observed in time-resolved
magnetization measurements has been interpreted in terms of a combination of a
Landau-Zener-St\"uckelberg transition
at zero field and spin-phonon coupling~\cite{Irinel1,Irinel4}.
Our results show that unless the field is applied in a special
direction ($x$ or $y$-direction in this case),
the adiabatic magnetization process is
no longer symmetric with respect to the field.
The dependence on the direction of the field should lead to
observable changes in the hysteresis loops.
So far, only weak directional dependence has been reported in experiments~\cite{private}.
Therefore it seems that it is necessary to explore other mechanisms that yield energy level repulsions such as
hyperfine interactions~\cite{HYPF}.

%
\begin{acknowledgments}
We thank 
I. Chiorescu, and V. Dobrovitski for illuminating discussions.
Support from the
`Nederlandse Stichting voor Nationale Computer Faciliteiten (NCF)'
is gratefully acknowledged.
The present work is partially supported by
Grant-in-Aid from the Ministry of Education, Culture, Sports,
Science and Technology, and also by NAREGI Nanoscience Project, Ministry of
Education, Culture, Sports, Science and Technology, Japan.
\end{acknowledgments}

\section*{Appendix A: Diagonalization of model (\ref{HD}) }\label{APPA}
Here we collect some analytical results of the solution of the eigenvalue problem
of Hamiltonian (\ref{HD}).
We only consider the case of a magnetic field that is parallel to the $z$-axis (${\bf h}=(0,0,h_z)$).
For a DM vector satisfying the conditions (\ref{d3sym}) and (\ref{c3sym})
the eight eigenvalues are given by

\begin{eqnarray}
E_{0,1}&=&\frac{-{\sqrt{3}}{D_z}  + 4h_z \pm {\sqrt{9{{D_x}}^2 + 9{{D_y}}^2 +
       (\sqrt{3}{D_z}  - 3J+2h_z)^2}}}{4},\nonumber \\
E_{2,3}&=&\frac{ -{\sqrt{3}}{D_z}  - 4h_z \pm   {\sqrt{9{{D_x}}^2 + 9{{D_y}}^2 +
       (\sqrt{3}{D_z}  - 3J-2h_z)^2}}}{4},\nonumber \\
E_{4,5}&=&\frac{  {\sqrt{3}}{D_z} \pm {\sqrt{3{{D_x}}^2 + 3{{D_y}}^2 +
       (\sqrt{3}{D_z}  + 3J-2h_z)^2}}}{4},\nonumber \\
E_{6,7}&=&\frac{  {\sqrt{3}}{D_z} \pm {\sqrt{3{{D_x}}^2 + 3{{D_y}}^2 +
       (\sqrt{3}{D_z}  + 3J+2h_z)^2}}}{4},\nonumber \\
\label{APPA0}
\end{eqnarray}
where $D_x=D^x_{1,2}$, $D_y=D^y_{1,2}$, and $D_z=D^z_{1,2}$.
Substituting the values of all model parameters,
we recover the results obtained by numerical diagonalization.
For $h_z=0$ there are four pairs of two-fold degenerate levels.
Although it is possible to find analytical expressions for the case that the
magnetic field is parallel to the $x$-axis, the expressions themselves are rather long and
not very illuminating. Therefore they are not given here.

For a magnetic field parallel to the $z$-axis,
a straightforward calculation shows that

\begin{eqnarray}
{\cal H}\left|{3}/{2},{3}/{2}\right\rangle&=&
-\frac{3(J+2h_z)}{4}\left|{3}/{2},{3}/{2}\right\rangle
+\frac{3(D_x+iD_y)}{4}\left|a\right\rangle,
\nonumber\\
{\cal H}^2\left|{3}/{2},{3}/{2}\right\rangle&=&
\frac{9[(J+2h_z)^2+D_x^2+D_y^2]}{16}\left|{3}/{2},{3}/{2}\right\rangle
-\frac{3(D_x+iD_y)(\sqrt{3}D_z+4h_z)}{8}\left|a\right\rangle,
\nonumber\\
{\cal H}\left|{3}/{2},{1}/{2}\right\rangle
&=&
-\frac{3J+2h_z}{4}\left|{3}/{2},{1}/{2}\right\rangle
-\frac{\sqrt{3}(D_x+iD_y)}{4}
\left|\bar a\right\rangle,
\nonumber\\
{\cal H}^2\left|{3}/{2},{1}/{2}\right\rangle
&=&
\frac{(3J+2h_z)^2+3(D_x^2+D_y^2)}{16}
\left|{3}/{2},{1}/{2}\right\rangle
-\frac{3(D_x+iD_y)D_z}{8}
\left|\bar a\right\rangle,
\label{APPA1}
\label{APPA2}
\end{eqnarray}
where $|\bar a\rangle$ denotes the state $|a\rangle$  with all spins reversed
and

\begin{eqnarray}
|a\rangle&=&\frac{1}{2\sqrt{3}}
\left[
(1-i\sqrt{3})|\downarrow\uparrow\uparrow\rangle
+(1+i\sqrt{3})|\uparrow\downarrow\uparrow\rangle
-2|\uparrow\uparrow\downarrow\rangle
\right].
\label{APPA3}
\end{eqnarray}
The expressions for
${\cal H}\left|{3}/{2},-{3}/{2}\right\rangle$,
${\cal H}^2\left|{3}/{2},-{3}/{2}\right\rangle$,
${\cal H}\left|{3}/{2},-{1}/{2}\right\rangle$, and
${\cal H}^2\left|{3}/{2},-{1}/{2}\right\rangle$
are obtained from Eqs.~(\ref{APPA1}) and (\ref{APPA3})
by changing the sign of $h_z$ and $D_y$ and replacing $|a\rangle$ by

\begin{eqnarray}
|b\rangle&=&\frac{-1}{2\sqrt{3}}
\left[
(1-i\sqrt{3})|\uparrow\downarrow\downarrow\rangle
+(1+i\sqrt{3})|\downarrow\uparrow\downarrow\rangle
-2|\downarrow\downarrow\uparrow\rangle
\right].
\label{APPA4}
\end{eqnarray}

Note that $\langle a|b\rangle=0$.
From Eq.~(\ref{APPA1}) it follows that
for the external field parallel to the $z$-axis,
model (\ref{HD}) does not allow transitions
from the state with all spins up (down) to the state with two spins down (up).
Therefore, if initially the system is in the state with all spins down,
adiabatically sweeping the field from a large negative value to a large positive value will
{\sl not} yield the final state with all spins up.

\section*{Appendix B: Numerical methods}\label{APPB}

A theoretical description of quantum dynamical phenomena in the \Mn\ and \V\ nanomagnets
requires detailed knowledge of their energy-level schemes.
Disregarding the fascinating physics of the nanomagnets, the calculation of the eigenvalues
of their model Hamiltonians is a challenging problem in its own right.
Firstly, the (adiabatic) quantum dynamics of these systems is mainly determined by the (tiny) level repulsions.
Therefore the calculation of the energy levels of these many-spin Hamiltonians has to be very accurate
in order to bridge the energy scales involved (e.g. from 500K to $\approx 10^{-9}$K).
Secondly, the level repulsions originate from the DM interactions that mix states with different magnetization.
In principle, this prevents the use of the magnetization as a vehicle to block-diagonalize the Hamiltonian
and effectively reduce the size of the matrices that have to be diagonalized.
If a level repulsion involves states of significantly different magnetization (e.g. $M^z=-10$ and $M^z=10$)
a perturbative calculation of the level splitting would require going to rather high order (at least 20),
a cumbersome procedure. Therefore it is of interest to explore alternative routes to
direct but accurate, brute-force diagonalization of the full model Hamiltonian.

As a non-trivial set of reference data, we used the eigenvalues obtained
by full diagonalization (using standard LAPACK algorithms)
of the $10000\times10000$ matrix representing model (\ref{MnHam}) \cite{Hans1}.
For one set of model parameters, such a calculation takes about 2 hours of CPU time on an
Athlon 1.8 GHz/1.5Gb system. Clearly this is too slow if we want to compute the energy-level diagram,
in particular if we want to estimate the structure of the level splittings.
At the resonant fields we need the eigenvalues
for many values of ${\bf h}$.
Furthermore, in the case of \V\ this calculation would take
about 30 times longer and require about 15 Gb of memory which, for present-day computers,
is too much to be of practical use.

We have tested different standard algorithms to compute the low-lying eigenvalues of large matrices.
The standard Lanczos method (including its conjugate gradient version)
as well as the power method~\cite{WILKINSON,GOLUB}
either converge too slowly, lack the accuracy to resolve the (nearly)-degenerate eigenvalues,
and sometimes even completely fail to correctly reproduce the low-lying part of the spectrum.
This is not a surprise: By construction these methods work well if the ground state is not degenerate
and there is little guarantee that they will work if there are (nearly)-degenerate
eigenvalues~\cite{WILKINSON,GOLUB}.
In particular, the Lanczos procedure suffers from numerical instabilities due to the loss of
orthogonalization of the Lanczos vectors~\cite{WILKINSON,GOLUB}.
It seems that model Hamiltonians for the nanoscale magnets provide a class of (complex Hermitian)
eigenvalue problems that are hard to solve.

Extensive tests lead us to the conclusion that only the Lanczos method with full orthogonalization
(LFO)~\cite{WILKINSON,GOLUB}
and the Chebyshev-polynomial-based projector method (CPP) discussed below can solve
these rather large and difficult eigenvalue problems with sufficient accuracy.
The former is significantly faster than the latter but using both gives
extra confidence in the results.

\subsection*{Lanczos method with full orthogonalization}
In the LFO, each time a new Lanczos vector is generated
we explicitly orthogonalize (to working precision) this vector to all, not just to the two previous,
Lanczos vectors~\cite{WILKINSON,GOLUB}.
With some minor modifications to restart the procedure when the Lanczos iteration terminates
prematurely, after $n$ steps this procedure transforms the $n\times n$ matrix $H$
into a tri-diagonal matrix that is comparable in accuracy
to the one obtained through Householder tri-diagonalization but offers no advantages~\cite{GOLUB}.
In our case we are only interested in a few low-lying eigenstates of $H$.
Thus we can exploit the fact that projection onto the (numerically exact) subspace of dimension $k$ ($k \ll n$)
built by the Lanczos vectors will yield increasingly accurate estimates
of the smallest (largest) eigenvalues and corresponding eigenvectors as $k$ increases.

In practice, to compute the $M$ lowest energy levels, the LFO procedure is carried out as follows.
\begin{itemize}
\item Perform a Lanczos step according to the standard procedure.
\item Use the modified Gramm-Schmidt procedure to orthogonalize the new Lanczos vector with respect to all
previous ones~\cite{WILKINSON,GOLUB}.
\item Compute the matrix elements of the tri-diagonal matrix.
\item At regular intervals, diagonalize the tri-diagonal matrix, compute the
approximate eigenvectors $\varphi_i$,
$\mu_i=\langle\varphi_i|H|\varphi_i\rangle$ and
$\Delta^2_i=\langle\varphi_i|(H-\mu_i)^2|\varphi_i\rangle$ for $i=1,\ldots,M$, and check if all
$\Delta_i$ are smaller than a specified threshold. If so, terminate the procedure (the exact
eigenvalue $E_i$ closest to $\mu_i$ satisfies $\mu_i-\Delta \le E_i\le \mu_i+\Delta_i$).
If not, continue generating new Lanczos vectors, etc.
\end{itemize}

\subsection*{Chebyshev polynomial projector method}\label{APPC}
As an alternative to the LFO,
we have used a power method~\cite{WILKINSON,GOLUB} based
on the matrix exponential $e^{-tH}$~\cite{DeRaedt87}.
Writing the random vector $\Psi(0)$ in terms of the (unknown) eigenvectors
$\{\phi_i\}$ of $H$, we find

\begin{eqnarray}
\Psi(t)=e^{-tE_0} \left[
\phi_0\langle\phi_0|\Psi(0)\rangle
+e^{-t(E_1-E_0)}
\phi_1\langle\phi_1|\Psi(0)\rangle
+e^{-t(E_2-E_0)}
\phi_2\langle\phi_2|\Psi(0)\rangle
+\ldots\right],
\label{psit}
\end{eqnarray}
showing $\lim_{t\rightarrow\infty}\Psi(t)/\Vert \Psi(t)\Vert\propto \phi_0$ if
$\langle\phi_0|\Psi(0)\rangle\not=0$.
In this naive matrix-exponential version of the power method, convergence to the lowest eigenstate
is exponential in $t$ if $E_1>E_0$.

The case of degenerate ($E_0=E_1=...$) or very close ($E_0\approx E_1\approx...$) eigenvalues can be
solved rather easily by applying the projector to a subspace instead of a single vector,
in combination with diagonalization of $e^{tH}$ within this subspace~\cite{DeRaedt87}.
First we fix the dimension $k$ of the subspace by taking $k$ equal or larger than the desired number of distinct eigenvalues.
The projection parameter $t$ should be as large as possible but nevertheless
sufficiently small so that at least the first $k$ terms survive one projection step.
Then we generate a set of random initial vectors $\Psi_i(0)$ for $i=1,\ldots,k$
and set the projection count $n$ to zero.
We compute the $k$ lowest eigenstates by the following algorithm~\cite{DeRaedt87}

\begin{itemize}
\item Perform a projection step $\Psi_i((n+1)t)=e^{-tH}\Psi_i(nt)$ for $i=1,\ldots,k$.
\item Compute the $k\times k$ matrices.
$A=\langle\Psi_i((n+1)t)|e^{tH}|\Psi_i((n+1)t)\rangle=
\langle\Psi_i((n+1)t)|\Psi_i(nt)\rangle$ and
$B=\langle\Psi_i((n+1)t)|\Psi_i(n+1)t)\rangle$. Note that $A$ is Hermitian and $B$ is positive definite.
\item Determine the unitary transformation $U$ that
solves the $k\times k$ generalized eigenvalue problem $Ax=\lambda Bx$. Recall that $k$ is small.
\item Compute $\Psi_i'((n+1)t)=\sum_{j=1}^k U_{i,j}\Psi_j((n+1)t)$ for $i=1,\ldots,k$.
\item Set $\Psi_i((n+1)t)=\Psi_i'((n+1)t)$ for $i=1,\ldots,k$.
\item Compute $\mu_i=\langle\Psi_i((n+1)t)|H|\Psi_i((n+1)t)\rangle$ and check if
$\Delta^2_i=\langle\Psi_i((n+1)t)|(H-\mu_i)^2|\Psi_i((n+1)t)\rangle$ is
smaller than a specified threshold for $i=1,\ldots,k$. If yes, terminate the calculation. If no, increase
$n$ by one and repeat the procedure.
\end{itemize}

We calculate $e^{-tH}\Psi$ by using the Chebyshev polynomial expansion
method~\cite{TAL-EZER,Iitaka01,LEFOR,SILVER,SlavaCheb}.
First we compute an upper bound $R$ of the spectral radius of $H$ (i.e., $\Vert H \Vert\le R$) by
repeatedly using the triangle inequality~\cite{SlavaCheb}.
From this point on we use the ``normalized'' matrix $\widetilde H = (2H/R-1)/2$.
The eigenvalues of the Hermitian matrix $\widetilde H$ are real and lie in the interval $[-1,1]$~\cite{WILKINSON,GOLUB}.
Expanding the initial value $\Psi(0)$ in the (unknown) eigenvectors $\phi_j$ of $\widetilde H$ (or $H$)
we find
\begin{equation}
\Psi(t)=e^{-tH}\Psi(0)=
e^{z\widetilde H}\Psi(0) =\sum_j e^{z\widetilde E_j} \phi_j
\langle\phi_j|\Psi(0)\rangle,
\label{expz}
\end{equation}
where $z=-tR$.
We find the Chebyshev polynomial expansion of $\Psi(t)$ by computing the
Fourier coefficients of the function $e^{z\cos\theta}$~\cite{ABRAMOWITZ}.
Alternatively, since $-1\le \widetilde E_j \le 1$, we can use the expansion
$
e^{z\widetilde E_j}=I_0(z) + 2\sum_{m=1}^{\infty} I_{m}(z)T_{m}(\widetilde E_j)
$
where $I_m(z)$ is the modified Bessel function of integer order $m$~\cite{ABRAMOWITZ}
to write Eq.~(\ref{expz}) as
\begin{equation}
\Psi(t)
=\left[I_0(z)I + 2\sum_{m=1}^{\infty} I_{m}(z) T_{m}(\widetilde H)\right] \Psi(0)\,.
\label{SUM0}
\end{equation}
Here, $I$ is the identity matrix and $T_{m}(\widetilde H)$ is the
matrix-valued Chebyshev polynomial defined by the recursion
relations
\begin{equation}
 T_{0}(\widetilde H)\Psi(0)=\Psi(0)\label{CHEB1}\,,\quad
 T_{1}(\widetilde H)\Psi(0)=\widetilde H\Psi(0)\,,
\label{CHEB44}
\end{equation}
and
\begin{equation}
 T_{m+1}(\widetilde H)\Psi(0)=
2\widetilde H T_{m}(\widetilde H)\Psi(0)- T_{m-1}(\widetilde H)\Psi(0)\,,
\label{CHEB4}
\end{equation}
for $m\ge1$.
In practice we will sum only contributions with $m\leq M$ where $M$ is chosen such that
for all $m>M$, $|I_m(z)/I_0(z)|$ is zero to machine precision.
Then it is not difficult to show that
$\Vert e^{-tH}/I_0(z) - I - 2\sum_{m=1}^{M} [I_{m}(z)/I_0(z)] T_{m}(\widetilde H)\Vert$
is zero to machine precision too (instead of $e^{-tH}$ we can equally well
use $e^{-tH}/I_0(z)$ as the projector).

Using the downward recursion relation of the modified Bessel functions,
we can compute $K$ Bessel functions to machine precision using
of the order of $K$ arithmetic operations~\cite{ABRAMOWITZ,NumericalRecipes}.
A calculation of the first 20000 modified Bessel functions takes less than 1 second
on a Pentium III 600 MHz mobile processor, using 14-15 digit arithmetic.
Hence this part of a calculation is a negligible fraction of the total computational work
for solving the eigenvalue problem.
Performing one projection step with $e^{-tH}$ amounts to repeatedly using recursion (\ref{CHEB4}) to
obtain $\widetilde T_{m}(B)\Psi(0)$ for $k=2,\ldots,M$, multiply the elements
of this vector by $I_{m}(z)$ and add all contributions.

\newpage

\end{document}